\documentclass[useAMS,usenatbib]{mn2e}
\usepackage{epsfig}
\usepackage{amsmath,amssymb}

\newcommand{\etal}{et al.\ }
\newcommand{\etalb}{et al.}

\def\lsim{~\rlap{$<$}{\lower 1.0ex\hbox{$\sim$}}}
\def\gsim{~\rlap{$>$}{\lower 1.0ex\hbox{$\sim$}}}
\def\psim{~\rlap{$\propto$}{\lower 1.0ex\hbox{$\sim$}}}

\title[Triple black hole systems]{Dynamics of triple black hole systems in hierarchically merging massive galaxies}

\author[Loren Hoffman and Abraham Loeb]{Loren Hoffman$^{1}$\thanks{E-mail:lhoffman@cfa.harvard.edu} and Abraham Loeb$^{2}$\thanks{E-mail:aloeb@cfa.harvard.edu}\\
$^{1}$Physics Department, Harvard University, 17 Oxford St.,
Cambridge, MA 02138, USA\\
$^{2}$Astronomy Department, Harvard University, 60 Garden Street, Cambridge, MA 02138, USA}

\begin{document}
\date{\today}
\pagerange{\pageref{firstpage}--\pageref{lastpage}} \pubyear{2006}
\maketitle
\label{firstpage}

\begin{abstract}
Galaxies with stellar bulges are generically observed to host supermassive
black holes (SMBHs). The hierarchical merging of galaxies should therefore
lead to the formation of SMBH binaries. Merging of old massive galaxies
with little gas promotes the formation of low-density nuclei where SMBH
binaries are expected to survive over long times.  If the binary lifetime
exceeds the typical time between mergers, then triple black hole (BH) systems
may form.  We study the statistics of close triple-SMBH encounters in 
galactic nuclei by computing a series of 3-body orbits with 
physically-motivated initial conditions appropriate for giant elliptical 
galaxies.  Our simulations include a smooth background potential consisting 
of a stellar bulge plus a dark matter halo, drag forces due to 
gravitational radiation and dynamical friction on the stars
and dark matter, and a simple model of the time evolution of the inner
density profile under heating and mass ejection by the SMBHs.  We find that
the binary pair coalesces as a result of repeated close encounters in
$\sim$85\% of our runs, and in $\sim$15\% of cases a new eccentric binary
forms from the third SMBH and binary remnant and coalesces during the run
time.  In about 40\% of the runs the lightest BH is left wandering through
the galactic halo or escapes the galaxy altogether, but escape of all three
SMBHs is exceedingly rare.  The triple systems typically scour out cores
with mass deficits $\sim$1-2 $\times$ their total mass, which can help to
account for the large cores observed in some massive elliptical galaxies,
such as M87.  The high coalescence rate, prevalence of very
high-eccentricity orbits, and gravitational radiation ``spikes'' during
close encounters in our runs, may provide interesting signals for the
future Laser Interferometer Space Antenna (LISA).
\end{abstract}


\setlength{\abovedisplayskip}{3pt}
\setlength{\belowdisplayskip}{3pt}
\setlength{\parskip}{0pt}

\section{Introduction}

In the favored cold dark matter cosmology, present-day galaxies were 
assembled hierarchically from smaller building blocks at earlier cosmic 
times.  Since all nearby galaxies with stellar spheroids are observed to 
host nuclear SMBHs \citep{KG01}, hierarchical merging leads inevitably to 
the formation of SMBH binaries \citep{BBR80}.  If the binary lifetime 
exceeds the typical time between mergers, then some galactic nuclei should 
contain systems of three or more SMBHs.  These systems are particularly 
interesting as they often lead to the ejection of one of the BHs 
at a speed comparable to the galactic escape velocity \citep{Loren}.  
In massive elliptical galaxies the typical speeds are 
$\sim 10^3~{\rm km~s^{-1}}$, far greater than attainable through 
gravitational radiation recoil \citep{Cent,FHH04,Will}.

Spatially resolved pairs of nuclei have been observed in a few active
galaxies.  The most famous example is NGC 6240, an Ultraluminous Infrared
Galaxy (ULIRG) in which two distinct active galactic nuclei (AGN) are
clearly seen in hard X-rays at a projected separation of $\sim$1 kpc
\citep{K03b}.  \citet{M95,M05} observed a variable UV source, possibly a
second active nucleus, at a projected separation of $\sim 60$ pc from the
primary nucleus in the spiral galaxy NGC 4736, which shows signs of a
recent merger.  \citet{R06} have detected what is thought to be an SMBH
binary at a projected separation of just 7.3 pc in the radio galaxy 0402+379,
through multi-frequency radio observations using the Very Long Baseline
Array (VLBA).  We begin by discussing the theory
of how such systems evolve, and the conditions under which they might
acquire a third BH.

\subsection{Black hole binaries}

When two galaxies merge, their dense nuclei sink to the center of the
merger product by dynamical friction.  As the nuclei spiral in, tidal
forces gradually strip the two SMBHs of their surrounding stars and dark
matter.  In mergers between galaxies of comparable mass, the BHs
are able to come together and form a bound SMBH binary on a timescale of
order $10^{9}$ yrs.  The binary continues to harden by dynamical friction
until it reaches a separation of order
\begin{equation}\label{ahard}
a_{hard} \equiv \frac{G\mu}{4\sigma^{2}} \approx 0.80 \frac{4q}{(1+q)^{2}} \left (\frac{m_{bin}}{10^{8} \hspace{2pt} {\rm M}_{\odot}} \right)^{1/2} \hspace{3pt} {\rm pc},
\end{equation}
known as the ``hardening radius'' (e.g. \citealt{Q96}).  Here $\mu =
m_{1}m_{2}/(m_{1}+m_{2})$ is the reduced mass of the two BHs with masses
$m_1$ and $m_2$, $\sigma$ is the velocity dispersion of the stars beyond
the binary's sphere of gravitational influence, $q$ is the binary mass
ratio $m_{2}/m_{1} \leq 1$, and $m_{bin}=m_{1}+m_{2}$ is the total mass of
the binary.  For smaller separations the binary looks like a point mass to
the distant stars contributing to dynamical friction, but close stellar
encounters preferentially harden the binary and so dominate further energy 
loss.  Only stars on nearly radial orbits, with periapsis distances of
order the binary separation, can extract energy from (``harden'') the
binary in this stage.  These stars undergo strong 3-body interactions 
with the binary and escape its vicinity with speeds comparable to the black
holes' orbital speed.  In the low-density nuclei of large elliptical
galaxies, the total mass in stars on such ``loss cone'' orbits is small
compared to the mass of the binary.  Furthermore the two-body stellar
relaxation time is long compared to a Hubble time, so once the stars
initially on loss cone orbits are cleared out, the loss cone remains empty
\citep{FR76,LS77,CK78}.  Since the binary must eject of order its own mass 
per $e$-folding in its semi-major axis, the system stops hardening around
$a_{hard}$ unless some other mechanism causes sufficient mass flux through
the binary.

If the binary reaches a separation around 
\begin{align}\label{agw}
a_{gw} = 4.5 &\times 10^{-2} \left( \frac{m_{bin}}{10^{8} \hspace{2pt} {\rm M}_{\odot}} \right)^{3/4} \left[ \frac{4q}{(1+q)^{2}} \right]^{1/4} \hspace{2pt} \cdot \notag \\
&\left( \frac{\tau_{gw}}{10^{10} \hspace{2pt} {\rm yrs}} \right)^{1/4} f^{-1/4}(e) \hspace{6pt} {\rm pc},
\end{align}
where $e$ is the orbital eccentricity of the binary and 
$f(e)=(1-e^{2})^{7/2}/(1+73e^{2}/24+37e^{4}/96)$, 
then it can coalesce on a timescale $\tau_{gw}$ through gravitational 
radiation \citep{BBR80}. To get from $a_{hard}$ to $a_{gw}$ it must bridge 
a gap
\begin{equation}\label{gap}
\frac{a_{hard}}{a_{gw}} \approx 17 \left( \frac{m_{bin}}{10^{8} \hspace{2pt} {\rm M}_{\odot}} \right)^{-1/4} \left[ \frac{4q}{(1+q)^{2}} \right]^{3/4} 
\end{equation}
by some mechanism other than stellar-dynamical friction or gravitational
radiation.  The question of whether and how it crosses this gap has become
known as the ``final parsec problem'' \citep{MMrev}.

In many galaxies there probably are alternative mechanisms for crossing the
gap.  When gas-rich galaxies merge, tidal torques channel large amounts of
gas into the central $\sim 100$ pc \citep{BSV87,H89}.  The gas may lose
energy through radiation and angular momentum through viscous torques, and
is therefore not subject to a loss cone problem.  Using Smoothed Particle
Hydrodynamics simulations \citet{ES04,ES05} compute a merger time of order
$10^{7}$ yrs in an environment typical of the central regions of ULIRGs,
which are thought to be gas-rich galaxies caught in the act of merging
\citep{S88}.  The nuclei of galaxies are also observed to contain
numerous massive perturbers (MPs) such as star clusters, molecular clouds, 
and possibly intermediate-mass black holes (IMBHs).  These objects scatter 
stars into the loss cone much more efficiently than other stellar mass 
objects, since the relaxation rate scales as the perturber mass for a fixed 
mass density of perturbers.  \citet{PHA06} extended the Fokker-Planck loss 
cone formalism \citep{FR76,LS77,CK78} to accomodate a spectrum of perturber 
masses and account for
relaxation by rare close encounters with MPs.  They show
that the population of known MPs in the nucleus of the Milky
Way is sufficient to bring a $4 \times 10^{6}$ M$_{\odot}$ BH binary to
$a_{gw}$ in $\sim 6 \times 10^{8}$ yrs, and it is reasonable to expect
similar perturber populations in other star-forming spiral galaxies.

The final parsec problem is often mentioned as a caveat when predicting the
SMBH coalescence signal in low-frequency gravitational wave detectors such
as the upcoming Laser Interferometer Space Anntena (LISA).  However the
LISA event rate is expected to be dominated by small galaxies at high
redshift \citep{WL03a,S05,RW05}, where the gas content and central
densities tend to be high and the relaxation times short.  For this reason
{\it the stalling problem is probably not a significant concern for the
LISA SMBH coalescence signal}.  On the other hand BH ejections by
gravitational radiation recoil \citep{M04,Haiman} may play an
important role in the high-redshift coalescence rate.  The long-term
survival of SMBH binaries is likewise unlikely in the gas-rich cores of
quasars and ULIRGs.

However none of the gap-crossing mechanisms discussed so far are likely to 
reduce the coalescence time below a Hubble time in mergers between 
giant, gas-poor elliptical galaxies.  \citet{MP04} show that a significant 
fraction of stars on ``centrophilic'' orbits in a triaxial potential can 
greatly increase the mass flux into the loss cone.  
Some non-axisymmetric potentials can also excite bar instabilities that 
cause rapid mass flow through the binary and efficient coalescence 
\citep{B06}.  However a central SMBH can disrupt box orbits and induce 
axisymmetry in the inner regions of a triaxial galaxy \citep{MQ98,H02}, 
and it is uncertain how often these geometry-specific mechanisms bring the 
coalescence time below a Hubble time.

One can naively assess the likelihood of coalescence by considering the 
``full'' and ``empty'' loss cone hardening times, $\tau_{full}$ and 
$\tau_{empty}$, in the nuclei of various galaxies assuming a spherical and
isotropic distribution function.  $\tau_{full}$ is the hardening time 
assuming every star kicked out of the loss cone is instantly
replaced, while $\tau_{empty}$ is the time assuming stellar two-body 
relaxation to be the only replenishing mechanism.  In small, dense 
galaxies $\tau_{full} \sim 10^{5-6}$ yrs and $\tau_{empty} \sim 10^{9-10}$ 
yrs while in the lowest-density cores of giant ellipticals and cD galaxies 
$\tau_{full} \sim 10^{8}$ yrs and $\tau_{empty} \sim 10^{14}$ yrs 
\citep{Y02}.  While the empty loss cone rate is difficult to believe in any 
galaxy given at least some clustering on scales larger than 1 M$_{\odot}$, it also seems
difficult to approach the full loss cone rate if there is no gas around and
no strong radial bias in the stellar distribution.  From this point of view
the stalling of binaries seems unlikely in small galaxies but probable in
low-density, gas-poor ellipticals.
If some binaries do survive for around a Hubble time, then the hierarchical 
buildup of galaxies will inevitably place three or more SMBHs in some 
merging systems.

\subsection{Merger-induced binary evolution before 3-body interactions: Back-of-the-envelope calculations}

An inspiralling satellite affects the evolution of a binary SMBH even long 
before it sinks to the center, by perturbing the large-scale potential and 
scattering stars into the loss cone.  We may estimate the extent of this 
effect as a function of satellite mass and distance from the center of the 
host galaxy using a rough but simple argument due to \citet{R81}.  The 
change in velocity necessary to deflect a star at radius $q$ into the loss 
cone is $\Delta V \sim h_{lc}/q$, where 
$h_{lc} \sim \sigma \sqrt{r_{inf}r_{bin}}$ is the characteristic 
specific angular momentum of stars on loss cone orbits \citep{FR76}, 
$r_{inf}=Gm_{bin}/\sigma^{2}$ is the SMBHs' radius of influence, and 
$r_{bin}$ is the binary separation.  The dynamical time at this radius is 
$t_{dyn} \sim q/\sigma$, so the acceleration required to scatter a star 
into the loss cone is roughly 
$a_{lc}(q) \sim \Delta V/t_{dyn} \sim \sigma^{2} \sqrt{r_{inf}r_{bin}}/q^{2}$.  Equating this with the tidal acceleration caused by the satellite,
$a_{tid}=2GM_{sat}(r)q/r^{3}$, where $r$ is the satellite's radius, yields
$q^{3}=\sigma^{2} \sqrt{r_{inf}r_{bin}}r^{3}/2GM_{sat}(r)$, or with
$r_{bin}=a_{hard}=G\mu/4\sigma^{2}$,
\begin{equation}\label{qsat}
q = \left[ \frac{\sqrt{m_{bin} \mu}}{4M_{sat}(r)} \right]^{1/3} r.
\end{equation} 
The $r$-dependence of $M_{sat}$ reflects the tidal stripping of the satellite
as it spirals inward.  Equation~\eqref{qsat} defines a critical radius $q$,
outside of which the satellite can deflect stars into (and out of) the loss 
cone in one dynamical time.  The mass flux through the binary induced by the
satellite is then approximately
\begin{equation}\label{lcflux}
\frac{dM_{stars}}{dt} (q) = 2\pi \rho(q) q^{2} \sigma \theta_{lc}^{2},
\end{equation}
where $\rho(q)$ is the density of the host galaxy at radius $q$ and 
$\theta_{lc}^{2} \approx r_{inf}r_{bin}/r^{2}$ is the geometrical factor
accounting for the fraction of stars on loss cone orbits as a function of 
radius $r$, assuming an isotropic distribution function \citep{FR76}.  For
a fixed satellite mass and distance, we can then define a ``binary feeding''
timescale by
\begin{equation}\label{tfeed}
\tau_{feed}=\frac{m_{bin}}{dM_{stars}/dt}=\frac{m_{bin}}{2\pi \rho(q) q^{2} \sigma \theta_{lc}^{2}(q)}.
\end{equation}

To determine whether the scattering of stars into the loss cone by the
satellite is sufficient to harden the binary enough to prevent a close
3-body encounter before the intruder arrives at the galactic center, we
must compare $\tau_{feed}$ with the timescale on which the satellite
spirals in by dynamical friction.  In the approximation of slow inspiral we
may write 
the dynamical friction timescale as $\tau_{df} \equiv |r/\dot{r}|\approx
|v/(dv/dt)_{df}|$.  Substituting $(dv/dt)_{df}$ from Chandrasekhar's
formula (equation~\eqref{df} in \S2.4.1; \citealt{C43}) yields
\begin{equation}\label{tdf}
\tau_{df}(r) = \frac{v_{sat}^{3}(r)}{4\pi G^{2} \rho(r) M_{sat}(r) [{\rm erf}(X)-\frac{2X}{\sqrt{\pi}}e^{-X^{2}}]},
\end{equation}
where $X \equiv v_{sat}/\sqrt{2}\sigma$.  $v_{sat}(r)$ in 
equation~\eqref{tdf} is computed from $v_{sat}(r)=\sqrt{GM_{host}(r)/r}$, 
where $M_{host}(r)=M_{stars}(r)+M_{halo}(r)+m_{bin}$ is the mass of the host
galaxy enclosed within radius $r$.  $M_{sat}(r)$ is the satellite mass 
contained within the tidal truncation radius obtained from a simple 
point mass approximation, $r_{tid}=[M_{sat}/M_{host}]^{1/3} r$ (this 
slightly underestimates $r_{tid}$ as the satellite approaches the center of 
the host).  In Fig. 1 we plot $\tau_{df}$ and $\tau_{feed}$ as a function 
of $r$ for a satellite with one third the stellar mass of the host, which 
contains a binary with 
$(m_{1},m_{2})=(1.2,3.7) \times 10^{8}$ M$_{\odot}$.  Both host
and satellite are modelled as Hernquist profiles \citep{H90}, with their 
masses and effective radii set by observed scaling relations. The details 
of the galactic model are described further in \S2.1-2.

\begin{figure}
\includegraphics[width=240pt,height=220pt]{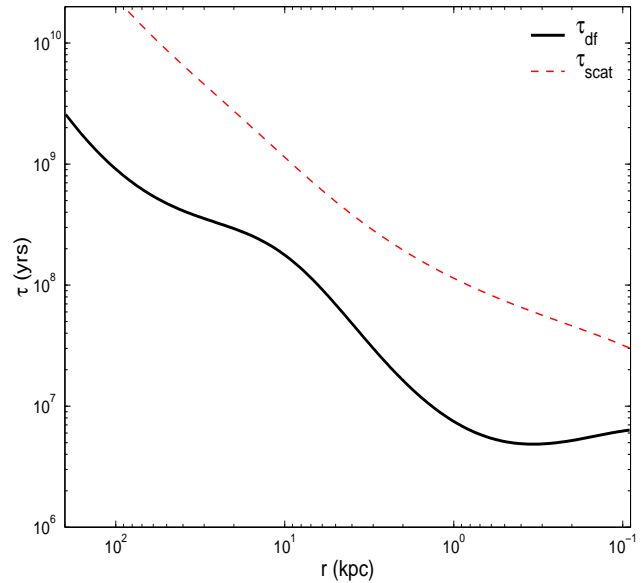}
\caption{Comparison of the ``feeding timescale'', $\tau_{feed}$, on which an 
inspiralling satellite scatters mass into the loss cone of an SMBH binary,
with the dynamical friction timescale, $\tau_{df}$, on which the satellite 
spirals in.  Upper (red dashed) line: $\tau_{feed}$ computed from 
equation~\eqref{tfeed}; Lower curve: $\tau_{df}$ computed from 
equation~\eqref{tdf}.  Both timescales are plotted as a function of the
satellite's distance from the center of the host galaxy.  For this plot we 
chose a binary mass of $4.5 \times 10^{8}$ M$_{\odot}$ and merger mass ratio
of 3:1 in the stars.  The galactic model is discussed in the text.}
\end{figure}

Since $\tau_{feed}$ remains about an order of magnitude above $\tau_{df}$
throughout the inspiral, this simple calculation makes it plausible that
the binary survives the merger process and undergoes close triple 
interactions with the infalling SMBH.  The tidal approximation (as well as 
our treatment of dynamical friction) breaks down as the satellite 
approaches $r_{inf}$, so the plot is cut off at a separation of 
$\sim 100$ pc, when the satellite still has $\sim$4 $e$-foldings to go to 
reach $a_{hard}$.  However this final stage of the inspiral is found to 
proceed very rapidly in N-body simulations \citep{QH97,MM01,M06}.  The 
merger's effect on the binary may be dominated by violent relaxation or
collective effects such as a bar instability \citep{B06}, in which case our 
two-body approach does not capture its essence.  The evolution of the core 
distribution function under the influence of a major merger is an 
intriguing open problem for simulators.

After the third BH becomes bound to the binary (but still before
the onset of close 3-body interactions) another hardening mechanism may
become important.  If the angle of inclination $i$ of the outer binary
(formed by the intruder and the inner binary center-of-mass (COM)) exceeds a
critical angle $\theta_{crit} \approx 39^{\circ}$, then the quadrupolar
perturbation from the intruder induces eccentricity oscillations through a
maximum \citep{K62}
\begin{equation}
e_{max} \approx \sqrt{1-\frac{5}{3} \cos^{2} i} .
\end{equation}
Since the gravitational radiation rate increases sharply toward high
eccentricities, these ``Kozai oscillations'' can greatly enhance the
radiation, possibly causing the binary to coalesce before it can undergo
strong 3-body interactions with the intruder \citep{BLS02}.  General
relativistic precession can destroy the Kozai resonance (e.g. \citealt{H97}),
but \citet{BLS02} find that this does not happen for 
\begin{equation}\label{KozCrit}
\frac{a_{out}}{a_{in}} \lsim 43 \left[ 2q_{out} \left( \frac{a_{in}}{1 \hspace{2pt} {\rm pc}} \right) \left( \frac{10^{8} \hspace{2pt} {\rm M}_{\odot}}{m_{bin}} \right) \right]^{1/3} \sqrt{\frac{1-e_{in}^{2}}{1-e_{out}^{2}}},
\end{equation}
where $a_{in}$ and $a_{out}$ are the semi-major axes of the inner and outer
binaries, $q_{out}$ is the outer binary mass ratio, and $e_{in}$ and 
$e_{out}$ are the inner and outer eccentricities. This leaves a window 
of about a factor of 10 in $a_{out}/a_{in}$ in which the Kozai mechanism
can operate before unstable 3-body interactions begin.

The actual enhancement of the gravitational radiation rate of course depends
on the amount of time spent at high eccentricity, but one may place an upper
limit on the importance of Kozai oscillations by computing the radiation
timescale if the inner binary spends all of its time at $e_{max}$.  The 
orbit-averaged power radiated by gravitational radiation is given by
\begin{equation}\label{gwpower}
\left|\frac{dE}{dt}\right|_{gw}=\frac{32G^{4}m_{1}^{2}m_{2}^{2}(m_{1}+m_{2})}{5c^{5}a^{5}} \frac{1+\frac{73}{24}e^{2}+\frac{37}{96}e^{4}}{(1-e^{2})^{7/2}}
\end{equation}
\citep{P64}, where $a$ is the semi-major axis and $e$ is the eccentricity.  
In Fig. 2 we plot contours of the gravitational radiation time 
$\tau_{gw}=|E/(dE/dt)_{gw}|$
in the $a$-$i$ plane by putting $e_{max}$ into equation~\eqref{gwpower},
for an equal-mass $6 \times 10^{8}$ M$_{\odot}$ binary.  
\begin{figure}
\includegraphics[width=240pt,height=220pt]{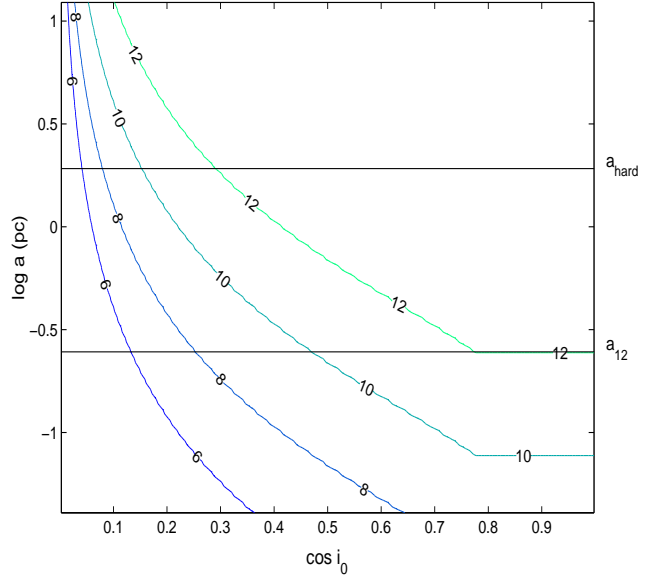}
\caption{Upper limit on the importance of Kozai oscillations in enhancing gravitational radiation by the inner binary ($m_{1}=m_{2}=3 \times 10^{8}$ M$_{\odot}$).  The cosine of the initial inclination angle is plotted on the horizontal axis, and the inner binary semi-major axis is plotted on the vertical axis.  Contours are plotted for gravitational radiation timescales of
$\tau_{gw} = 10^{12}$, $10^{10}$, $10^{8}$, and $10^{6}$ yrs, if the binary 
were to stay at maximum eccentricity throughout the whole oscillation cycle. 
The horizontal lines indicate the hardening radius and the separation such
that a circular binary would coalesce on a $10^{12}$ yr timescale.}
\end{figure}
This may seem like a gross overestimate of the gravitational radiation rate, 
especially since the shape of the Kozai oscillations is in fact such that 
the binary spends more time near $e_{min}$ than near $e_{max}$.  
However since $\tau_{gw}$ is so strongly dominated by periapsis passages 
at $e \approx e_{max}$, the shift in the contours for a realistic high-$e$ 
duty cycle is only modest.  See \citep{BLS02} for comparison with a detailed 
study of radiation enhancement by Kozai oscillations in binaries with 
initial $\tau_{gw} \sim 10^{12}$ yrs.  For a binary at $a_{hard}$, Kozai 
oscillations can induce coalescence within $10^{10}$ yrs in $\lsim 20$\% 
of cases assuming $\cos i$ is uniformly distributed.  In the remainder of 
cases the inner binary may survive until the outer binary shrinks to the 
point of unstable 3-body interactions.

\subsection{Close 3-body encounters}

If the intruder comes close enough before it causes sufficient hardening of 
the (inner) binary, then a strong 3-body encounter takes place.  Strong 
encounters are characterized by a significant transfer of energy between 
the binary's internal degrees of 
freedom and the COM motion of the binary and third body.
When the intruder is slow relative to the binary's orbital speed $v_{bin}$, 
energy typically flows from the inner binary to the outer components, 
so that the binary is more strongly bound after the encounter.  This 
is one manifestation of the negative specific heat characteristic of 
gravitationally bound systems.  The encounter ends in the escape of one of 
the three bodies, usually the lightest, from the system at a speed 
comparable to $v_{bin}$. 

When the lightest body $m_{3}$ escapes, momentum conservation requires that
the binary COM recoil in the opposite direction
with a speed smaller by a factor $m_{3}/(m_{1}+m_{2})$.  It is instructive
to compare the expected ejection velocities of the binary and $m_{3}$ with 
the typical galactic escape velocity.  For a circular binary with 
$m_{1}=m_{2}=m_{bin}/2$, the binding energy at the hardening radius is
$E_{B, hard}=Gm_{bin}^{2}/8a_{hard} \approx 6.8 \times 10^{55} [m_{bin}/(10^{8} \hspace{3pt} {\rm M}_{\odot})]^{3/2}$ erg.
The binding energy at the radius where $\tau_{gw}=10^{8}$ yrs is
$E_{B, gw}=Gm_{bin}^{2}/8a_{gw} \approx 6.2 \times 10^{57} [m_{bin}/(10^{8} \hspace{3pt} {\rm M}_{\odot})]^{5/4}$ erg.  The mean energy
$\Delta E$ harvested from the binary in close encounters with slow intruders
is about $0.4E_{B}$, though the median $\Delta E$ is somewhat lower 
\citep{HF80}.  Energy conservation implies that the escaper leaves the
system with kinetic energy $KE_{sing}=\Delta E/[1+m_{3}/(m_{1}+m_{2})]$ 
while the binary leaves with $KE_{bin, cm}=\Delta E/[1+(m_{1}+m_{2})/m_{3}]$ 
in the system COM frame.  For an equal mass binary with 
$m_{bin}=5 \times 10^{8}$ M$_{\odot}$, this gives ejection velocities of
$v_{sing} \sim 290 {\rm km/s}$ and $v_{bin} \sim 140 {\rm km/s}$ for the
binary at $a_{hard}$, and $v_{sing} \sim 4000 {\rm km/s}$ and 
$v_{bin} \sim 2000 {\rm km/s}$ for the binary at $a_{gw}$.  

Any nonzero eccentricity of the binary will increase the semi-major axis 
corresponding to a fixed $\tau_{gw}$, lowering the ejection velocities for 
the binary at $a_{gw}$.  Also any deviation from equal masses will 
result in a smaller fraction of the extracted energy being apportioned to 
the binary and a smaller binary recoil velocity.  The typical escape 
velocity for galaxies hosting $5 \times 10^{8}$ M$_{\odot}$ BHs is around 
1500 ${\rm km/s}$, accounting for both the stars and the dark matter.  From 
these numbers, it appears that single escapes will be fairly common as 
repeated encounters harden the binary to $\sim a_{gw}$.  However accounting 
for realistic mass ratios and eccentricities (the first 
3-body encounter tends to thermalize the eccentricity even if it 
starts off 
circular), binary escapes should be rare.  Since the binary must come near 
the escape velocity to remain outside the nucleus for a significant amount 
of time, we do not expect triple interactions to empty many nuclei of BHs.  
We will quantify these statements with our triple-BH simulations.

The formation of triple SMBH systems through inspiral of a merging satellite
leads to a rather specific initial configuration. 
The three BHs start off as a bound 
``hierarchical triple,'' consisting of an inner binary with 
$a_{in} \sim a_{hard}$ and a more widely separated outer binary with
semi-major axis $a_{out}$.  
For very large $a_{out}/a_{in}$ we expect hierarchical triples to exhibit 
very regular behavior; in this case the third body sees the inner binary as 
a point mass and the system essentially consists of two independent (inner
and outer) binaries.  However as $a_{out}/a_{in}$ approaches unity, secular
evolution gives way to chaotic 3-body interactions in which the orbits 
diverge and the system becomes subject to escape of one its components.  
\citet{MA01} derive a criterion for the stability of 3-body systems 
based on an analogue with the problem of binary tides.  The most distant 
intruder orbit at which unstable interactions can begin is reliably 
estimated by  
\begin{equation}\label{MAstab}
\frac{R_{p}^{out}}{a_{in}} \approx 2.8 \left[ \frac{(1+q_{out})(1+e_{out})}{\sqrt{1-e_{out}}} \right]^{2/5},
\end{equation}
where $R_{p}^{out}$ is the periapsis separation of the outer binary, 
$a_{in}$ is the semimajor axis of the inner binary, 
$q_{out}=m_{3}/(m_{1}+m_{2})$ is the outer binary mass ratio, and
$e_{out}$ is its eccentricity.  This criterion has great practical importance
due to the high numerical cost of unnecessarily following weak 
hierarchical systems.  It specifies an optimal 
starting point for our simulations, which aim to study strong interactions 
in 3-body systems starting off as hierarchical triples.

Naively one might expect a strong 3-body encounter following a merger with 
a galaxy hosting a binary, so long as the intruder does
not induce coalescence of the binary before it reaches the center.  However
the stability criterion implies a condition for close interactions much more
stringent than this.  To undergo a chaotic encounter with the inner binary,
the intruder must reach the stability boundary before the outer binary 
hardens and 
stalls.  A triple system covers somewhat more stellar phase space than a
binary of the same size, but not by much for a stable hierarchical 
system.  This means that the merger process cannot cause the binary to 
harden by more than around an $e$-folding for a nearly circular, equal-mass
system before the intruder arrives at the 
center.  Though the order-of-magnitude estimates in the previous section
make this plausible, further study is needed to determine the likelihood of 
unstable triple interactions in realistic merger situations.  An eccentric 
outer binary 
relaxes the criterion somewhat, but dynamical friction tends to 
circularize the orbits of satellites with moderate initial eccentricities
before they reach the nucleus \citep{MM01}.  We therefore assume 
near-circular initial orbits and begin each simulation from a weakly 
hierarchical configuration.

\subsection{Previous work and goals of this study}

Triple SMBH systems in galactic nuclei were first considered by \citet{S74},
who computed an extensive series of Newtonian 3- and 4-body orbits, 
and compared the slingshot ejection statistics to the observed structure of 
extragalactic radio sources.  \citet{V76} included a gravitational radiation 
drag force in the 3-body dynamics.  He showed that this perturbing force 
could in some cases yield much higher ejection velocities than would be 
possible in Newtonian gravity, with associated bursts of gravitational 
waves.  

The more complex problem of three or four SMBHs coming together in
the hierarchical merging process and interacting in a galactic
potential was first addressed by \citet{MV90} and \citet{V94}, who 
experimented with a variety of initial BH configurations.  \citet{H01} 
studied binary-binary scattering in galactic nuclei using initial 
conditions (ICs) based on Extended Press Schechter theory \citep{LC93}.
\citet{V03a} followed the formation of triple BH systems in halo merger
trees tracking the hierarchical buildup of SMBHs from $\sim 150$ 
M$_{\odot}$ seeds in high-$\sigma$ peaks at $z \approx 20$.  Using a simple 
analytic prescription for the ejection velocities, they inferred the 
presence of a large population of SMBHs and IMBHs wandering through the
halos of galaxies and intergalactic space.  \citet{I05} performed the first 
full N-body simulations of equal-mass triple BH systems embedded in stellar 
bulges, an important contribution to our understanding of galactic nuclei.  
Because of the large computation time required for each run, they could not 
statistically sample the highly varied outcomes of the 3-body encounters as 
the previous authors did.  

In this paper we study the dynamics of repeated triple-SMBH interactions in
galactic nuclei.  Between close encounters we follow the wandering BHs
through the galaxy as their orbits decay by dynamical friction.  We use
physically-motivated initial BH configurations and mass distributions, and 
updated galactic models characteristic of the low-density, massive 
elliptical galaxies in which SMBH binaries are most likely stall.  
We include both
a stellar and a dark matter component, with the stellar spheroid fixed to
lie on the observed $m_{bh}-\sigma$ and $m_{bh}-M_{bulge}$ relations 
\citep{T02,FM00,MH03,M98}.  The close encounters are treated using a 
KS-regularized Bulirsch-Stoer integrator provided by Sverre Aarseth 
\citep{MA90,MA93}.  The inner density profile is updated throughout the 
simulations to roughly account for core heating by dynamical friction and 
stellar mass ejection.  Gravitational radiation losses are modelled as a 
drag force determined by the relative coordinates and velocities of each 
pair.  Each 
simulation takes only a few minutes to run, so we can try a variety of 
distributions of ICs and statistically sample the outcomes for each.
We use this algorithm to study a variety of consequences of the ongoing
encounters, such as the merging efficiency of BH pairs, the time spent
wandering at various distances from the galactic center, the distribution
of final sizes and eccentricities of the binaries remaining in the galaxy 
after a steady state has been reached, and the extent of the core scouring 
caused by the triple SMBH systems.

Aside from the motivating order-of-magnitude calculations in previous
sections, this paper does not address the question of {\it whether}
close triple SMBH systems form in galactic nuclei.  We start our simulations
from a state that the system must reach shortly before the onset of 
unstable 3-body interactions {\it assuming} that they occur, and 
proceed to derive the subsequent evolution.  Our results may be used to 
argue for or against the occurrence of triple systems in real galaxies, as 
observations support or disfavor the signatures that we derive.

In \S2 we describe our model and code methods.  In \S3 we present
the results of our study, and in \S4 we discuss these results and 
conclude.

\section{Model and Methods}

\subsection{BH mass distribution and halo model}

To get a physically motivated distribution of BH mass ratios, we 
associate the formation of the inner and outer binaries with the last two
major mergers in the history of the galactic halo hosting the triple system.
We use Extended Press-Schechter theory \citep{LC93} to calculate the 
probability distributions of the halo formation times and progenitor 
masses, and randomly select the parameters of the previous two mergers from 
these distributions.  We then assign a BH to each progenitor halo using a 
simple prescription based on the assumption of a flat galactic rotation 
curve.

\citet{LC93} derive the instantaneous halo merger rate,
\begin{multline}\label{eps}
r_{LC}(M_{1},M_{f},t) = \frac{d^{2}p}{dM_{2} dt} = \sqrt{\frac{2}{\pi}} \frac{\delta_{c}}{D(z)} \hspace{3pt} \left| \frac{\dot{\delta_{c}}}{\delta_{c}} - \frac{\dot{D}}{D} \right| \hspace{3pt} \cdot \\
\frac{\left| d\sigma/dM \right|_{M_{f}}}{\sigma^{2}(M_{f})} \hspace{5pt} \frac{\exp \left[-\frac{\delta_{c}^{2}}{2D^{2}(z)} \hspace{2pt} (\frac{1}{\sigma^{2}(M_{f})}-\frac{1}{\sigma^{2}(M_{1})})\right]}{[1-\sigma^{2}(M_{f})/\sigma^{2}(M_{1})]^{3/2}}.
\end{multline}
This equation gives the probability, per unit time per unit mass 
of $M_{2}$, of a given halo of mass $M_{1}$ merging with another halo of 
mass $M_{2}$ to form a product of mass $M_{f}$ $=$ $M_{1}+M_{2}$ at time 
$t$.  Here $\sigma^{2}(M)$ is the present-day variance of the linear 
density field on mass scale $M$,
\begin{equation}
\sigma^{2}(M) = \frac{1}{(2\pi)^{3}} \int_{0}^{\infty} P(k) W^{2}(kr) 4\pi k^{2} dk,
\end{equation}
where $P(k)$ is the power spectrum of density fluctuations today, $W$ is a 
tophat window function, and $r$ is related to $M$ through 
$M=(4/3)\pi r^{3} \rho_{m}$, the volume times the present-day matter 
density.  $P(k)$ is related to the primordial power spectrum through the
transfer function $T(k)$, which encapsulates the suppression of 
perturbations on small scales due to radiation pressure and damping over 
the history of the universe.  For $T(k)$ we adopt the standard fitting 
formulae of \citet{EH98}.  For the linear growth function $D(z)$ we use
the approximation
\begin{equation}
D(z) \approx \frac{(5/2) \hspace{2pt} \Omega_{m}(z) D_{0}(z)}{\Omega_{m}^{4/7}(z)-\Omega_{\Lambda}(z)+\left[ 1+\frac{\Omega_{m}(z)}{2}\right] \left[ 1+\frac{\Omega_{\Lambda}(z)}{70} \right]},
\end{equation}
good to within a few percent for all plausible values of 
$\Omega_{m}$ and $\Omega_{\Lambda}$ \citep{C92}.  $D_{0}(z) = 1/(1+z)$ is 
the growth function for an Einstein-de Sitter universe, 
$\Omega_{m}(z)=\Omega_{m}(1+z)^{3}/[\Omega_{m}(1+z)^{3}+\Omega_{\Lambda}]$ 
is the matter density 
(normalized to the critical density) as a function of redshift, and we take 
$\Omega_{\Lambda}(z)=1-\Omega_{m}(z)$ assuming the rest of the density is in
the form of a cosmological constant.  $\delta_{c}$ has the weak redshift
dependence \citep{KS96}
\begin{equation}
\delta_{c} \approx \frac{3(12\pi)^{2/3}}{20} [1+0.0123 \log_{10} \Omega_{m}(z)].
\end{equation}
We adopt the cosmological parameters obtained from three years of data
collection by the Wilkinson Microwave Anisotropy Probe (WMAP), 
$\Omega_{m} h^{2}=0.127$, $\Omega_{b} h^{2}=0.0223$, $h=0.73$, 
$\sigma_{8}=0.74$, and $n_{s}=0.951$ \citep{S06}.

Since the merger rate~\eqref{eps} diverges as $M_{2}/M_{1}$ $\rightarrow$ 0, 
applications of the formula that track individual merging halos must employ a
cutoff mass ratio $M_{2}/M_{1}$ $\equiv$ $\Delta_{m}$, such that all 
mergers below $\Delta_{m}$ are treated as smooth {\it accretion} 
rather than as discrete mergers (see \citealt{MPS96} for further discussion).
The instantaneous rate of accretion onto a halo of mass $M$ at redshift 
$z$ is
\begin{equation}\label{racc}
r_{a}(M,t) = \int_{M}^{M(1+\Delta_{m})} (M'-M) r_{LC}(M,M',t) dM'.
\end{equation}
To get the growth history (``accretion track'') of a halo of mass $M_{0}$ 
at time $t_{0}$ due to accretion since the last merger, one need only solve 
the differential equation $dM/dt$ $=$ $r_{a}[M(t),t]$, subject to the 
initial condition $M(t_{0})$ $=$ $M_{0}$.  We integrate this equation 
backward in time using a 4th-order Runge-Kutta method to get the accretion 
tracks of the halos in our simulations.  Since we are interested in BH 
binary formation, we loosely associate $\Delta_{m}$ with the halo mass 
ratio such that tidal stripping of the satellite would prevent the eventual 
merging of the two nuclei.  N-body simulations of galaxy mergers place this 
mass ratio in the range $\Delta_{m}$ $\sim$ $0.1-0.3$, depending on the 
density and orbital parameters of the satellite \citep{TEA03,CMG99}.  Hence 
our canonical choice is $\Delta_{m}$ $=$ 0.3, and we also try values of 
$\Delta_{m}=0.1$ (runs D1) and 0.5 (runs D5), the latter being the halo 
mass that corresponds to a stellar mass ratio of $\sim$ 3:1 in our 
prescription. 

Following \citet{MPS98}, we write the probability, per unit time, of a halo
with mass $M_{f}$ at time $t$ arising from a merger with a smaller halo of 
mass between $M$ and $M+dM$ (the ``capture rate'') as
\begin{multline}\label{rcap}
r_{c}(M,M_{f},t) \hspace{2pt} dM = \\
r_{LC}(M,M_{f},t) \theta[M_{f}-M(1+\Delta_{m})] \hspace{2pt} \frac{N(M,t)}{N(M_{f},t)} \hspace{2pt} dM,
\end{multline} 
the EPS merger rate excluding halos below the threshold 
$\Delta_{m}$, and weighted by the number of mass $M$ halos per unit halo of 
mass $M_{f}$.

The rate at which halos of mass $M_{f}$ form through all mergers at time 
$t$ is \begin{equation}\label{rform}
r_{f}(M_{f},t) \approx \frac{1}{2} \int_{M_{f}\Delta_{m}/(1+\Delta_{m})}^{M_{f}} r_{c}(M,M_{f},t) dM.
\end{equation}
The probability distribution function (PDF) of formation times of halos 
with mass $M_{0}$ at time $t_{0}$ is 
\begin{equation}\label{phif}
\Phi_{f}(M_{0},t) = r_{f}[M(t),t] \hspace{2pt} e^{-\int_{t}^{t_{0}} r_{f}[M(t'),t'] dt'}.
\end{equation}
Given a formation time $t_{f}$ and corresponding mass $M(t_{f})$ along the 
past accretion track of $M_{0}$, the mass of the larger progenitor $M_{1}$
is distributed according to
\begin{equation}\label{phip}
\Phi_{p}[M(t_{f}),M_{1}] = \frac{2G(M_{1},M)}{\int_{M\Delta_{m}/(1+\Delta_{m})}^{M/(1+\Delta_{m})} G(M',M) dM'},
\end{equation}
where
\begin{equation}\label{gfunc}
G(M',M) = \frac{|d\sigma(M')/dM'|}{M'\sigma^{2}(M')} \left[ 1-\frac{\sigma^{2}(M)}{\sigma^{2}(M')}  \right]^{-3/2}.
\end{equation}
By choosing formation times and progenitor masses randomly according to 
\eqref{phif} and \eqref{phip}, we capture the stochasticity of the 
intervals between mergers above $\Delta_{m}$, but treat merging below this 
threshold only in the mean.  See \citet{MPS01}, \citet{MPS98}, 
and \citet{MPS96} for further details and derivations 
of~\eqref{phif} and~\eqref{phip}. 
Fig. 3 shows the distribution of formation times, progenitor masses, and
accretion tracks for a present-day $5 \times 10^{13}$ M$_{\odot}$ halo for
$\Delta_{m}=$ 0.1, 0.3, and 0.5, and the accretion tracks for 
1, 2.3, and 5 $\times 10^{13}$ M$_{\odot}$ halos with $\Delta_{m}$ fixed at
0.3.  All accretion tracks are normalized to the present-day mass $M_{0}$.
Note the insensitivity of the shape of these tracks to $M_{0}$, as expected
for masses above the critical mass $M^{*} \approx 2 \times {\rm 10}^{12} {\rm M}_{\odot}$.
\begin{figure}
\includegraphics[width=250pt,height=230pt]{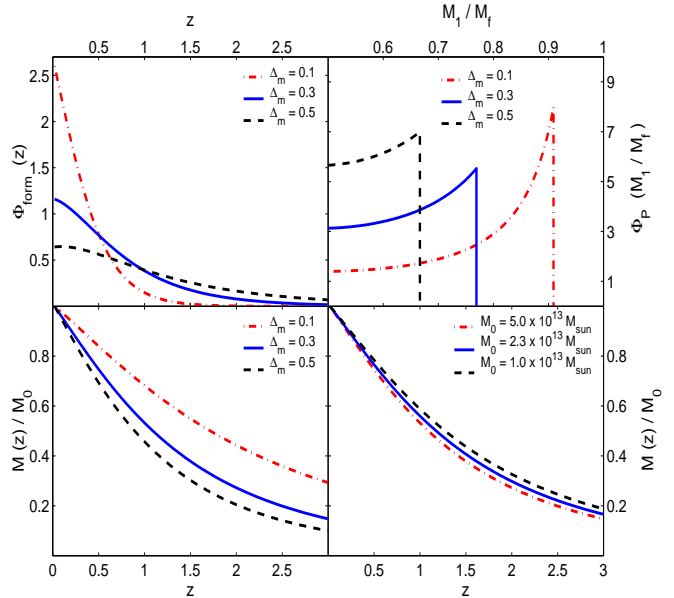}
\caption{{\it Upper left:} Probability distribution function (PDF) of formation redshifts given by equation~\eqref{phif} for a $5 \times 10^{13} M_{\odot}$ halo at $z=0$.  {\it Upper right:} PDF of
masses of the larger progenitor of the same halo given by equation~\eqref{phip}, normalized to the mass of the merger product, $M_{f}=5 \times 10^{13} M_{\odot}$.  {\it Lower left:} Past accretion tracks of a present-day $5 \times 10^{13} M_{\odot}$ halo back to $z=3$, normalized to the mass at $z=0$.  {\it Lower right:} Normalized accretion tracks for three different halo masses.}
\end{figure}

Our algorithm for generating the BH masses is illustrated schematically in
Fig. 4.  For each run we begin 
with a halo of mass $M_{0}$ $=$ $5 \times 10^{13}$ $M_{\odot}$ at time 
$t_{0}$ $\equiv$ $t(z=0)$, choose it's formation time $t_{f0}$ randomly 
according to equation \eqref{phif}, and find the mass $M_{f0}=M(t_{f0})$ 
along its accretion track at that time.  The mass $M_{f0}$ is assigned to 
the dark matter halo hosting the triple BH system, and the physical 
time for the run to end if other termination conditions are not met 
first is set to $t_{0}-t_{f0}$.  To explore the dependence of the results
on the absolute mass scale, we also try beginning with a 
$1 \times 10^{13}$ $M_{\odot}$ halo (runs H1).

We model the halo as a Hernquist 
profile \citep{H90}, which is identical to an NFW 
profile \citep{NFW} in its inner regions if the scale radius $a_{H}$ is 
related to the NFW scale radius by 
$a_{H}$ $=$ $a_{NFW} \sqrt{2[\log(1+c)-c/(1+c)]}$,
where $c$ is the halo concentration defined by $a_{NFW}=r_{vir}/c$.  The
Hernquist model falls off as $r^{-4}$ instead of $r^{-3}$ far outside
$a_{H}$ \citep{SDH05}.  The virial radius $r_{vir}(M,z)$ is given by
\begin{multline}\label{rvir}
r_{vir}(M,z) = \\
\frac{364}{1+z} \left[ \frac{M_{halo}}{10^{13} h^{-1} {\rm M}_{\odot}} \frac{\Omega_{m}(z)}{\Omega_{m}} \frac{18\pi^{2}}{\Delta_{c}} \right]^{1/3} h^{-1} {\rm kpc},
\end{multline}
where $\Delta_{c}$ $=$ 
$18 \pi^{2}+82[\Omega_{m}(z)-1]-39[\Omega_{m}(z)-1]^{2}$ 
\citep{BL01}, and $c$ roughly follows the median relation
from the $\Lambda$CDM simulations of \citet{B01}, $c \approx 9.0 [(2.1
\times 10^{13} {\rm M}_{\odot}) / M_{halo}]^{0.13}/(1+z)$.  The
$z$ dependence of $r_{vir}$ and $c$ nearly cancel to make $a_{H}$ depend
only weakly on redshift, so we simply use the $z=0$ relation between 
$M_{halo}$ and $a_{H}$ in our simulations.

We choose the mass $M_{1}$ of the larger progenitor of $M_{f0}$ 
randomly according to equation~\eqref{phip}, and assign a mass 
$M_{2}$ $=$ $M_{f0}-M_{1}$ to the smaller progenitor.  
Before the merger the larger progenitor is assumed to have hosted a BH 
binary, while the smaller one hosted a single BH.  Repeating 
the procedure used for $M_{0}$, we assign formation times $t_{f1}$ and 
$t_{f2}$ to $M_{1}$ and $M_{2}$ using equation~\eqref{phif}, and choose
progenitor masses $M_{11}$, $M_{12}$, $M_{21}$, and $M_{22}$ according to 
equation~\eqref{phip}.

Having constructed a set of progenitor halos, we now need a BH-halo relation
$m_{bh}(M_{halo},z)$ to complete our algorithm.  We obtain such a relation 
by equating the halo virial 
velocity $v_{vir}$ to the circular velocity $v_{c}$ of the stellar spheroid,
and using empirical $v_{c}-\sigma$ and $\sigma-m_{bh}$ correlations to 
connect $v_{c}$ to $m_{bh}$, similar to the approaches in 
\citet{EKB06} and \citet{WL05}.  Combining
\begin{multline}\label{vvir}
v_{vir} = {\rm 343} \hspace{5pt} \sqrt{1+z} \hspace{5pt} \times \\
\left( \frac{M_{halo}}{10^{13} h^{-1} {\rm M}_{\odot}} \right)^{1/3} \left[\frac{\Omega_m}{\Omega_{m}(z)} \frac{\Delta_{c}}{18\pi^{2}} \right]^{1/6}  {\rm km \hspace{3pt} s^{-1}} 
\end{multline}
\citep{BL01}
with $v_{c}$ $\approx$ $314 [\sigma / (208 \hspace{3pt} {\rm km/s})]^{0.84}$ ${\rm km/s}$ \citep{F02} and $\sigma/(208 \hspace{3pt} {\rm km/s})$ 
$\approx$ $[m_{bh}/(1.56 \times 10^{8} {\rm M}_{\odot})]^{1/4.02}$ 
\citep{T02}, we arrive at the relation
\begin{equation}\label{mmbh}
\left(\frac{M_{halo}}{10^{12} M_{\odot}} \right)= 8.28
\left(\frac{M_{bh}}{10^{8} M_{\odot}}\right)^{0.626} \gamma(z),
\end{equation}
where $\gamma(z) \equiv (1+z)^{-3/2} [(\Omega_{m}/\Omega_{m}(z))(\Delta_{c}/18\pi^{2})]^{-1/2}$.

In our canonical runs we set the masses of the inner binary members to
$m_{bh}(M_{11},z_{f1})$ and $m_{bh}(M_{12},z_{f1})$, and that of the
intruding BH to $m_{bh}(M_{21}+M_{22},z_{f2})$.  Note that in this
prescription the intruder is usually lighter than the heavier binary member,
so that most of the 3-body interactions
result in an exchange.  To examine the effect of more interactions without
exchange, we try choosing $m_{bh}[{\rm max}(M_{21},M_{22}),z_{f2}]$
for the intruder mass in runs (MX).  As there is neither a direct causal 
relationship
between $m_{bh}$ and $M_{halo}$ predicted by theory \citep{WL05} nor a
tight correlation directly observed between these two variables, and we
know that identical halos may host galaxies of different morphologies and
occupation numbers, $m_{bh}(M_{halo},z)$ should be taken with something of
a grain of salt. Nevertheless it is a useful way to generate simple but
physically-motivated BH mass distibutions when no information other
than the halo mass is available.  
\begin{figure}
\includegraphics[width=240pt,height=240pt]{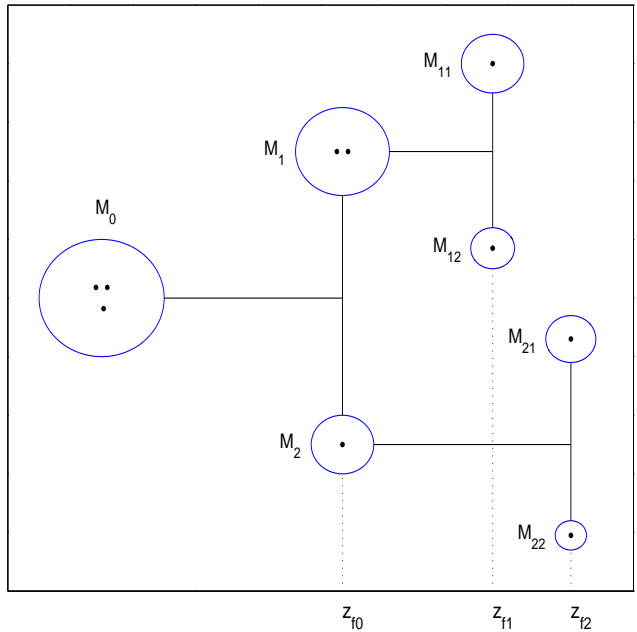}
\caption{Schematic diagram of our algorithm for generating the BH mass distribution.  First we select a formation time $z_{f0}$ for the halo $M_{0}$ 
hosting the BH triple system randomly from equation~\eqref{phif}.  Given $z_{f0}$ and $M(z_{f0})$, we select two progenitor masses $M_{1}$ and $M_{2}$ 
according to equation~\eqref{phip}, assign the binary to the larger one 
and the third BH to the smaller one.  We repeat this process going back
one step further in the ``merger tree'' to get the masses of the binary 
constituents.}
\end{figure}

We make one final modification to the set of BH masses
used in our simulations.  If the outer binary's hardening radius lies 
outside the stability boundary given by equation~\eqref{MAstab} with 
$a_{in}=a_{hard}$, then the decay of the outer 
orbit is expected to stall before a strong encounter can begin.  
To roughly account for this we exclude all ICs where 
$\mu_{out}$ $>$ $3\mu_{in}$. 
The final distribution of BH mass ratios is shown in Fig. 5 for 
$\Delta_{m}=$0.1, 0.3, and 0.5.  
\begin{figure}
\includegraphics[width=240pt,height=215pt]{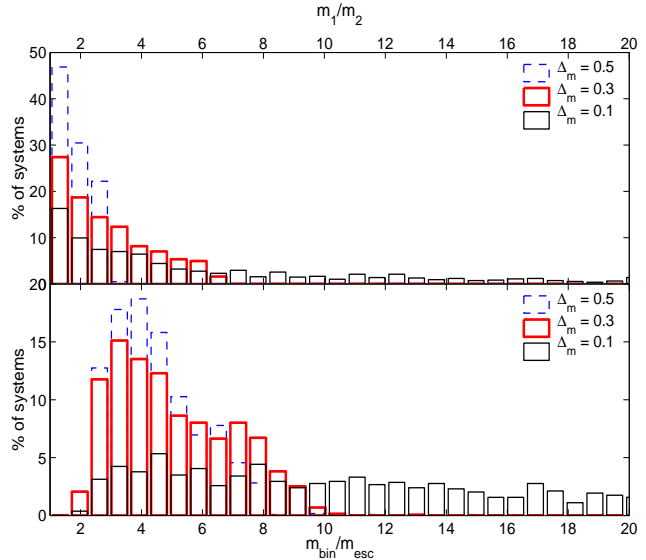}
\caption{Distribution of BH binary mass ratios.  {\it Upper panel:} Inner binary mass ratio $m_{1}/m_{2} > 1$.  {\it Lower panel:} $m_{bin}/m_{esc}$, where $m_{esc}$
is the mass of the lightest BH and $m_{bin}$ is the sum of the masses of the
other two BHs.}
\end{figure}
In the upper panel we plot the inner 
binary mass ratios, while the lower panel shows the distribution of 
$m_{bin}/m_{esc}$, where $m_{esc}$ is the mass of the lightest BH and 
$m_{bin}$ is the sum of the masses of the other two BHs.  This ratio
determines the binary recoil speed when the lightest BH is ejected from the
system.  The total BH mass is typically $\sim 6 \times 10^{8}$ M$_{\odot}$
in our canonical runs.   

\subsection{Stellar spheroid model}

To complete the galactic model we surround the BH system by a smooth stellar
potential superimposed on the dark matter halo.  The stars are modelled using
the ``$\eta$-models'' of \citet{T94}, with a sharp break to shallower 
slope $-\gamma$ added at $r_{b} \ll a$:
\begin{equation}\label{rhor}
\rho(r) = 
\begin{cases}
{\displaystyle \frac{\eta}{4\pi} \frac{Ma}{r^{3-\eta}(r+a)^{1+\eta}} \equiv \rho_{\eta}(r)},  &\text{if $r>r_{b}$;} \vspace{3pt} \\ 
{\displaystyle \rho_{\eta}(r_{b}) (r/r_{b})^{-\gamma}} &\text{if $r<r_{b}$.}
\end{cases}
\end{equation}
Our canonical model is the $\eta=2$ (Hernquist) profile, and we also try 
$\eta=1.5$ (runs SC) to explore the effect of a steeper inner profile and
higher central density ($\rho \sim 800 {\rm \hspace{2pt} M_{\odot} \hspace 
{3pt}pc^{-3}}$ for $\eta=1.5$ vs. 
$\rho \sim 180 {\rm \hspace{2pt} M_{\odot} \hspace 
{3pt}pc^{-3}}$ for the Hernquist profile at the BH radius of influence).  
$r_{b}$ and $\gamma$ were initialized to reflect
the cusp destruction caused by the inspiralling BHs in reaching their
initial configuration, and were updated throughout the simulation to
account for the continued core heating and mass ejection.  Our algorithm
for updating the core is described further in \S3.5.

The parameters $M$ and $a$ in the $\eta$-models were set based on the 
tight correlations observed between SMBH mass $m_{bh}$ and stellar 
bulge mass \citep{M98,MH03,P06} and velocity dispersion 
\citep{FM00,G00,T02}.  \citet{MH03} found the relation 
$M_{MH03}=({\rm 4.06} \times {\rm 10}^{10} M_{\odot})(m_{bh}/10^{8} M_{\odot})^{1.04}$ between $m_{bh}$ and the 
virial mass $M_{vir} = k R_{e} \sigma_{e}^{2}/G$ of the stellar bulge,
where $R_{e}$ is the half-light radius and $\sigma_{e}$ is the effective
bulge velocity dispersion.  They set $k=3$ ($k$ would be 8/3 for an 
isothermal sphere) to get an average ratio of unity between $M_{vir}$ and 
the dynamically measured masses $M_{dyn}$ of galaxies with more direct 
stellar-dynamical mass determinations \citep{G03}.
$\sigma_{e}$ is typically measured over either a circular aperture of radius
$R_{e}/8$ \citep{FM00} or a linear aperture out to $R_{e}$ \citep{G00} --
\citet{T02} discuss the essential agreement between the velocity 
dispersions measured in these two ways.  Thus for each model ($\eta=2,1.5$)
we compute the projected radius $R_{e} \equiv \kappa_{1} a$ containing 
half the integrated surface brightness (assuming a constant mass-to-light
ratio), and velocity dispersion 
$\sigma_{e}^{2} \equiv \kappa_{2}GM/a$ at radius $R_{e}/8$.  Here 
$\kappa_{1}$ and $\kappa_{2}$ are constants depending on
the density profile (see \citealt{T94} for relevant formulae).  The 
parameter $M$ is then chosen to satisfy 
$3 R_{e} \sigma_{e}^{2}/G = 3\kappa_{1}\kappa_{2}M = M_{MH03}(m_{bh})$ 
$\Rightarrow$ $M = M_{MH03}/(3\kappa_{1}\kappa_{2})$.  For the 
Hernquist model with $\kappa_{1}=1.815$ and $\kappa_{2}=0.104$,
$M_{H}=1.76M_{MH03}=(7.15 \times 10^{10} M_{\odot})(m_{bh}/10^{8} M_{\odot})^{1.04}$.  The scale radius is then obtained from $a=GM_{MH03}/3\kappa_{1}\sigma_{bh}^{2}(m_{bh})$, where $\sigma_{bh}(m_{bh})$ is the velocity dispersion computed from the $m_{bh}-\sigma$ relation of \citet{T02}. In each of these 
relations $m_{bh}$ is set to the total mass of the triple BH system.

In a perfectly smooth, spherically symmetric galactic
potential, BHs ejected on distant radial orbits return 
directly to the center to interact strongly with any other nuclear black
holes.  Since real galaxies are clumpy and triaxial, the interaction will
more realistically be delayed until the orbit of the ejected BH 
decays by dynamical friction.  To mitigate this problem we flattened
the $\eta$-models by adding two low-order spherical harmonic terms to the
spherical potential \citep{dZC96}:
\begin{equation}\label{triax}
V(r)=u(r)-v(r)Y_{2}^{0}(\theta)+w(r)Y_{2}^{2}(\theta,\phi),
\end{equation}
where $u(r)$ is the potential of the spherical $\eta$-model,
$v(r)=-GMr_{1}r^{\eta-1}/(r+r_{2})^{\eta+1}$, and
$w(r)=-GMr_{3}r^{\eta-1}/(r+r_{4})^{\eta+1}$.  Since
near-sphericity in the inner regions is probably a necessary prerequisite
for the survival of the inner binary for of order a Hubble time until the 
next merger \citep{MP04,B06}, the parameters
$r_{1},...,r_{4}$ were chosen to give a spherical profile near the galactic
center, and axis ratios approaching 1.3 and 1.5 for $r \gg a$.  A similar
triaxial modification was applied to the dark matter halo, and the
relative orientation of the stellar and halo potentials was chosen
randomly.  By misaligning their axes we eliminate any artificial stable
orbits (e.g. along the long axis of an ellipsoid) near which ejected BHs
tend to return on a perfectly radial orbit to the center.  This triaxial
modification had the desired effect of preventing frequent strong
encounters at periapsis on distant orbits, but had little influence on the
global outcome statistics.

\subsection{Initial BH configuration}

We assume that the 3-BH system starts off as a hierarchical triple on
the verge of unstable 3-body interactions.  In our canonical runs we 
initialize the inner binary semi-major axis $a_{in}$ to $a_{hard}$.
To study the effect of varying $a_{in}$ we also try runs with
$a_{in}=3r_{h}$ (runs BA) and $a_{in}=r_{h}/3$ (runs SA).  The outer binary 
semi-major axis $a_{out}$ is set by the stability criterion of \citet{MA01},
equation~\eqref{MAstab}.  The initial eccentricity of
the inner (outer) binary was chosen uniformly between 0.0 and 0.2
(0.3), in accordance with the low eccentricities found in galaxy merger
simulations where dynamical friction tends to circularize the orbits of 
satellites as they spiral inward \citep{MM01}.  
The three Euler angles of the intruder's orbital 
plane were chosen randomly relative to the reference plane of the binary 
orbit, as was the phase of the initial periapsis of the binary. Both 
orbits were always started at periapsis; since many orbital periods elapse 
before unstable interactions begin, the relative phase is effectively 
randomized in any case.  Having defined an initial configuration of three 
BHs embedded in a stellar+dark matter potential, we next describe how we
evolve the system forward in time.

\subsection{Code method}

We treat the close 3-body encounters using Sverre Aarseth's {\it Chain} 
code, an implementation of the N-body regularization technique of Mikkola 
and Aarseth \citep{MA90,MA93}.  The masses are first ordered
so that neighbors in the chain are the dominant two-body
interactions, then the KS-transformation \citep{KS65} is applied to 
neighboring pairs.  This transformation eliminates the singularity at
$r \rightarrow 0$ in Newtonian gravity and transforms the equations of 
Keplerian motion to the simple harmonic oscillator equation \citep{SS71}.  
External perturbing forces of arbitrary strength depending on the 
coordinates, velocities, and/or time are simply incorporated into the 
formulation (though of course singularities in these perturbing forces need 
not be eliminated by the change of variables).  We use this to add a 
galactic potential (\S2.1-2), a gravitational radiation back-reaction 
force, and a stellar-dynamical friction force on the intruding BH.
The regularized equations of motion are integrated using the
Bulirsch-Stoer (BS) method \citep{BS66} based on Romberg extrapolation.
For unperturbed sinusoidal motion, the BS integrator requires only two or 
three timesteps per orbital period!  

When the binary and third body are far apart we switch to two-body 
motion (of the single BH and binary COM) using a 4th-order Runge-Kutta (RK4) 
method.  We simultaneously evolve the binary semi-major axis and eccentricity
using orbit averaged equations, $da = [(da/dt)_{st}+(da/dt)_{gw}] dt$ and 
$de = [(de/dt)_{gw}] dt$, where $(d/dt)_{st}$ and $(d/dt)_{gw}$ are the 
contributions from stellar interactions and gravitational radiation. The 
timesteps are adaptively controlled with a simple step-doubling
scheme: at each step the 14 numbers 
$\{x_{1},...,x_{6};v_{1},...,v_{6};a,e\}$ are all required to remain the
same to within an error $\epsilon$ $=$ $10^{-n}$ under doubling of the step
size.  To avoid wasting computation time when any of these values approach
zero, we accept agreement to $n$ decimal places as an alternative criterion
for convergence.  For the calculations reported in this paper we set
$n=12$.

The relative perturbation to the binary from the third body at apoapsis, 
\begin{equation}\label{pert}
\delta F = \frac{2 r_{ap}^{3} m_{3}}{\min(m_{1},m_{2}) d^{3}},
\end{equation}
is used to decide which integration method to use at any given time.  Here
$r_{ap}$ is the apoapsis distance between $m_{1}$ and $m_{2}$, $m_{3}$ is
the intruder mass, and $d$ is the distance of the intruder from the binary
COM.  We switch to two-body RK4 integration each time $\delta F$ 
falls below $5 \times 10^{-5}$ and call the {\it Chain} code again when 
$\delta F$ reaches $5 \times 10^{-4}$.  We choose different $\delta F$ 
thresholds for beginning and ending close encounters to prevent overly 
frequent toggling between the two methods.  When $<3$ BHs remain in 
the simulation (after coalescence of the inner binary or escape of one or 
more BHs from the galaxy), we primarily use the RK4 integrator, but 
call the {\it Chain} code to treat very close two-body encounters.  Since 
chain regularization is defined only for three or more bodies, we add a 
light and distant ``dummy'' particle when using this method for two-body 
motion. 

During the two-body motion we declare the single BH or binary (remnant) to
have escaped if its distance from the galactic center exceeds 500 kpc and
its specific energy $E = \Phi(r,\theta,\phi) + \frac{1}{2}v^{2}$ 
exceeds $E_{esc}$, the energy needed to escape from $r=0$ to infinity.
We declare the binary to have coalesced during a close encounter when
(i) $r_{12} < 3r_{sb}$, where $r_{sb}$ is the Schwarzchild radius of the 
larger member of the pair, or (ii) $|a/\dot{a}|_{gr}$ $<$ $0.1t_{dyn}$ and 
$|a/\dot{a}|_{gr}$ $<$ 50 yrs while $\delta F < 10^{-3}$, where $t_{dyn}$ is 
the current outer binary dynamical time.  During the RK4 integration we
require that $|a/\dot{a}|_{gr}$ $<$ 50 yrs or $r_{12} < 1.1(r_{s1}+r_{s2})$
at periapsis, where $r_{s1,2}$ are the Schwarzchild radii of the two 
binary members.  Upon coalescence we replace the pair with a single body of 
mass $m_{bin}$ and the COM position and velocity.
A run ends when (a) only one SMBH remains in the galaxy and it 
has settled to the center of the potential by dynamical friction; (b) two
BHs remain and have formed a hard binary at the galactic center; (c) the 
physical time exceeds $t_{max} = t_{0}-t_{f0}$, the current age of the 
universe minus the halo formation time; (d) all BHs have escaped the galaxy;
or (e) the physical time spent in a call to {\it Chain} exceeds a maximum 
allowed time $t_{chn}$.  The last condition is added to avoid spending too 
much computation time on very long close encounters. 

\subsubsection{Treatment of stellar-dynamical friction}

During the two-body evolution we apply a dynamical friction force given by
Chandrasekhar's formula \citep{C43,BT87},
\begin{equation}\label{df}
\left( \frac{d\vec{v}}{dt} \right)_{df} = - \frac{4\pi G^{2} \rho m \ln \Lambda [{\rm erf}(X)-2X e^{-X^{2}}/\sqrt{\pi}]}{v^{2}} \hat{v},
\end{equation}
where $X \equiv v/(\sqrt{2} \sigma)$, to the single BH and binary COM.
The factor in square brackets $\approx$ $1$ for $v \gg \sigma$ and 
$\approx$ $0.75X^{3}$ for $v \ll \sigma$.  We take
\begin{equation}\label{logL}
\ln \Lambda = {\rm max} \left\{ \ln \left[ \frac{r (\sigma^{2}+v^{2})}{Gm} \right], 1 \right\}
\end{equation}
for the Coulomb logarithm, where $r$ is the BH's distance from the galactic 
center.  For $\rho$ in equation~\eqref{df} we use 
${\rm min}[\rho(r),\rho(r_{inf})]$, effectively capping the density at its
value at the BH radius of influence, $r_{inf}=Gm/\sigma^{2}$, when the BHs
pass through the core.

The semi-major axis $a$ of the binary also evolves under stellar-dynamical 
friction as it wanders through the galaxy.  However Chandrasekhar's formula 
applied separately to the binary constituents does not give a good 
description of this evolution, since the hard binary loses energy through 
close 3-body encounters with stars, while equation~\eqref{df} 
relies on the assumption that the energy loss is dominated by weak 
two-body encounters.  We approximate the evolution of $a$ using a 
formulation for the decay rate of a hard, massive binary in a uniform and 
isotropic sea of stars developed in \citet{MV92} and \citet{Q96}.  The 
formulation was calibrated with an extensive series of 3-body 
scattering experiments in \citet{Q96} and tested against N-body simulations 
in \citet{MV92}.  The binary decay rate is given by
\begin{equation}\label{dadtQ}
\left( \frac{da}{dt} \right)_{st} = - \frac{\rho H a^{2}}{\sigma}, 
\end{equation}
where the hardening rate $H$ can be approximated by the empirical fitting 
function \citep{Q96}
\begin{equation}\label{Hq96}
H \approx \frac{16}{\left[ 1+ \left(\sigma/w \right)^{4} \right]^{1/2}}.
\end{equation} 
Here $w=0.85\sqrt{G \hspace{3pt} {\rm min}(m_{1},m_{2})/a}$ is the 
characteristic velocity distinguishing the hard binary regime -- stars with
$v \gsim w$ cannot be easily captured into bound orbits and preferentially
harden the binary in close encounters.
In our simulations the binary COM is often speeding
through the stellar medium at $v_{cm} \gsim \sigma$ after an energetic 
ejection, so the stellar medium looks ``hotter'' in its frame of 
reference.  To account for this we replace $\sigma$ in
equations~\eqref{dadtQ} and~\eqref{Hq96} with
$\sigma^{*} \equiv \sqrt{v_{cm}^{2}+\sigma^{2}}$,
a good approximation since $H$ is not very sensitive to the shape of the 
distribution function (e.g. $H \approx 16$ for a Maxwellian vs. 
$H \approx 18$ for a uniform velocity distribution).  For $\rho$ in 
equation~\eqref{dadtQ} we took ${\rm min}[\rho(r),\rho(r_{inf})]$ as we did
for the drag on the COM.  We ignored the mild eccentricity 
evolution $(de/dt)_{st}$, which is shown in \citet{Q96} to be far weaker 
than that predicted by Chandrasekhar's formula for hard eccentric binaries.

When the amplitude of oscillation of one of the two masses falls below
$Gm/2\sigma^{2}$, we stop integrating its motion and 
place it at rest at the galactic center until the second body returns to
within a distance of twice the break radius, $2r_{b}$.  
If the settled mass is the binary, then we also stop 
updating its semi-major axis for stellar hardening, assuming that it clears 
out its loss cone and stalls once it stops moving 
about the nucleus and encountering new stars.  Since the total mass in 
loss cone stars is small compared to the BH mass
in the low-density galaxies that we consider, to good approximation the 
binary stalls as soon as the replenishing mechanism (motion) shuts off.

During close encounters between the three BHs an orbit-averaged
prescription for stellar-dynamical friction is not feasible.  However the 
triples are still marginally stable at the boundary given by 
equation~\eqref{MAstab}, so we apply a drag force given by Chandrasekhar's
formula with $\ln \Lambda = 1$ to the intruder at the beginning of each run. 
At the onset of chaotic interactions in the first encounter (defined 
loosely by the first time the closest pair is not formed by the original 
binary members) this perturbation is shut off, and it remains off in all
later close encounters.  Fortunately the chaotic interactions occur on 
timescales very short compared to a dynamical friction time, so it is valid 
to neglect stellar dissipation during close encounters.

\subsubsection{Treatment of gravitational radiation}

Gravitational radiation is modelled using the 
${\cal O}[(v/c)^{5}]$ post-Newtonian (2.5PN) back-reaction acceleration 
computed by \citet{DD81}, evaluated in the two-body 
COM frame (e.g. Gultekin \etal 2006),
\begin{multline}\label{gwdd}
\frac{d\vec{v_{1}}}{dt} = \frac{4G^{2}}{5c^{5}} \frac{m_{2}}{m_{1}+m_{2}} \frac{m_{1}m_{2}}{r^{3}} {\Big \{} \hat{r} (\hat{r} \cdot \vec{v}) {\Big [} \frac{34G(m_{1}+m_{2})}{3r} \\
+6v^2 {\Big ]} + \vec{v} {\Big [} -\frac{6G(m_{1}+m_{2})}{r}-2v^{2} {\Big ]} {\Big \}}.
\end{multline}
$\vec{r}=\vec{r}_{1}-\vec{r}_{2}$ and
$\vec{v}=\vec{v}_{1}-\vec{v}_{2}$ are the relative positions and velocities
of the two masses.  We sum the force linearly over all pairs, a valid
approximation provided the perturbations from the third body and other
external tidal forces are instantaneously small at periapsis.  When
averaged over a complete orbit, equation~\eqref{gwdd} is equivalent to the
\citet{P64} equations for the binary semi-major axis and eccentricity,
\begin{subequations}\label{pet}
\begin{align}
\frac{da}{dt} &= -\frac{64}{5} \frac{G^{3}m_{1}m_{2}(m_{1}+m_{2})}{c^{5}a^{3}} \frac{1+\frac{73}{24}e^2+\frac{37}{96}e^{4}}{(1-e^{2})^{7/2}}\label{pet1} \\
\frac{de}{dt} &= -\frac{304}{15} \frac{G^3m_{1}m_{2}(m_{1}+m_{2})}{c^{5}a^{4}} \frac{e+\frac{121}{304}e^{3}}{(1-e^{2})^{5/2}}.\label{pet2}
\end{align}
\end{subequations}
However when $|\hat{v} \cdot \hat{r}|$ comes close to one on hyperbolic 
orbits, so that $(\hat{r} \cdot \vec{v})^{2} \rightarrow v^{2}$, 
$\dot{E} = \vec{F}_{1} \cdot \vec{v}_{1} + \vec{F}_{2} \cdot \vec{v}_{2}=m_{1} \vec{a}_{1} \cdot \vec{v}$ as given by equation~\eqref{gwdd}
becomes {\it positive}, though we know physically that gravitational waves
can only carry energy away from the system.  To give the correct answer 
averaged over an orbit, this positive contribution must be cancelled by 
extra energy loss near periapsis, making the equation potentially 
sensitive to numerical error.  This effect is much less pronounced in the 
\citet{DD81} form than in other expressions derived for the 
radiation back-reaction acceleration - they derived the formula specifically
for practical use on the problem of two point masses (see Appendix of 
\citealt{L93} and references therein).

For computational ease we neglect the lower-order 1-2PN terms (precession 
of the periapsis) in the post-Newtonian expansion.  Though much larger in
magnitude than the radiation reaction force, these terms are unimportant in
the statistical sense because they conserve the intrinsic properties of the
system, such as energy \citep{K06,I05}.  We need not concern ourselves with
relativistic precession destroying the Kozai resonance since the semi-major
axis ratio given by equation~\eqref{MAstab} is much smaller than that of
equation~\eqref{KozCrit}.

\subsection{Code tests and energy errors}

One way to establish the reliability of our integration methods is to test 
them on problems with known solutions.  Fig. 6 shows an example on the 
two-body problem with gravitational radiation.  
\begin{figure}
\includegraphics[width=250pt,height=275pt]{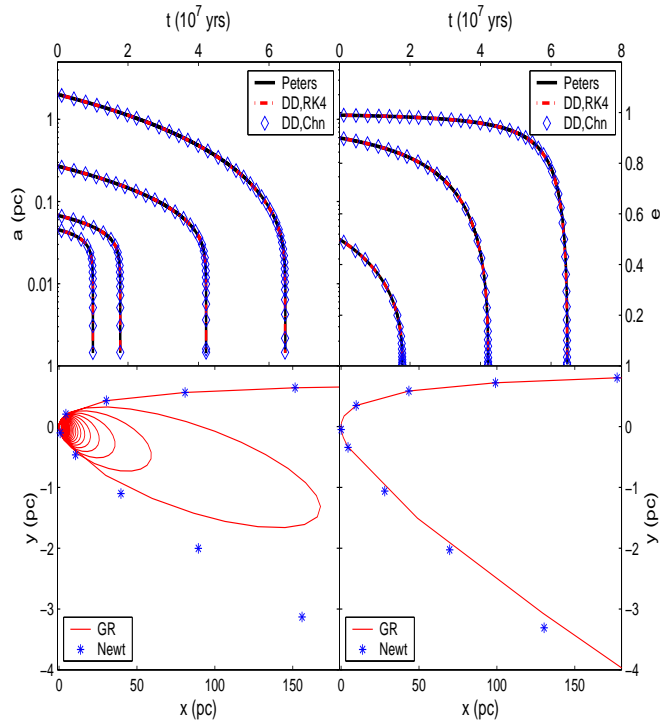}
\caption{Code tests on the Newtonian two-body problem with gravitational radiation.  {\it Upper panels:} Evolution of elliptical orbits computed using RK4 integration of the Peters equations (black solid), RK4 integration with the Damour \& Deruelle (DD) radiation back-reaction acceleration (red dashed), and {\it Chain} integration with the DD acceleration (diamonds).  The ICs were chosen to give $|a/\dot{a}|$ $=$ $10^{7}$ yrs at the beginning of each integration.  {\it Left:} Semi-major axis evolution for initial eccentricities of $e_{0}$ $=$ 0.0, 0.5, 0.9, and 0.99 (bottom to top).  {\it Right:} Eccentricity evolution for $e_{0}$ $=$ 0.5, 0.9, and 0.99.  The curves' indistinguishability demonstrates the reliability of all three methods.  {\it Lower panels:} Hyperbolic orbits with impact parameters $b$ set to 80\% and 120\% of the critical value for gravitational radiation capture, computed using the DD acceleration in the {\it Chain} code.  {\it Left:} 0.8$b_{crit}$; BH is captured.  {\it Right:} 1.2$b_{crit}$; BH is not captured.  The blue asterisks are points along the Newtonian trajectory (without gravitational radiation).  The deviation from the Newtonian trajectory after periapsis can be seen in both plots, even though the energy remains positive in the latter.}
\end{figure}
The upper panels show the
evolution of the semi-major axis $a$ and eccentricity $e$ of four decaying
elliptical orbits, computed using (a) our RK4 integrator and 
equation~\eqref{gwdd}, with an error tolerance of $\epsilon=10^{-9}$, 
(b) the {\it Chain} code and equation~\eqref{gwdd}, with 
$\epsilon=10^{-14}$, and (c) the \cite{P64} equations~\eqref{pet}.  In 
each case the initial semi-major axis $a_{0}$ was chosen to give a 
gravitational radiation timescale of $|a/\dot{a}| \approx 10^{7}$ yrs, and 
the four curves (from bottom to top) are for eccentricities of 0.0, 0.5, 
0.9, and 0.99.  The agreement of the three computation methods demonstrates 
the reliability of both the RK4 integrator and our implementation of the 
{\it Chain} code in handling dissipative forces. 

The lower panels show two hyperbolic orbits with periapsis distances around
30 times the Schwarzchild radius $r_{sb}$ of the larger BH, computed using 
equation~\eqref{gwdd} in {\it Chain}.  The RK4 integrator was found to 
fail some tests on very close approaches from hyperbolic orbits with 
gravitational radiation, so we treat all such approaches using the 
regularized {\it Chain} code in our runs, even during the unperturbed 
binary evolution.  The blue asterisks are points along the Newtonian 
orbits while the red solid lines show the trajectories with gravitational
radiation.  There is a simple analytic expression for the maximum
periapsis distance for gravitational radiation capture from a hyperbolic
orbit,
\begin{equation}\label{bcrit}
r_{p,max}=\left[ \frac{85 \sqrt{2} \pi G^{7/2} m_{1}m_{2}(m_{1}+m_{2})^{3/2}}{12 c^{5} v_{\infty}^{2}} \right]^{2/7},
\end{equation}
where $m_{1}$ and $m_{2}$ are the masses of the two bodies and $v_{\infty}$
is their relative velocity at infinity.  The orbit on the lower left
begins at 80\% of the critical impact parameter and the incoming BH is 
captured.  On the right the intruding BH starts at 120\% of the critical
impact parameter and is not captured, though the deviation from the Newtonian
trajectory due to the energy radiated at periapsis can be seen on the way
out.  We tried iterating over impact parameters close to the critical value
and found that the code reproduces equation~\eqref{bcrit} to within a part 
in $10^{6}$ for periapsis distances $r_{peri} \sim 30r_{s}$, and to within 
a part in $10^{3}$ for $r_{peri} \sim 3r_{s}$.

We also evaluated the performance of the code by repeating our canonical
set of 1005 runs with a static
inner profile to check the precision of energy conservation.  In Fig. 7 we
histogram the energy errors, computed as
\begin{equation}\label{epsE}
\epsilon = \left| \frac{E_{0} + \sum_{i} \left[ \int_{t_{0}}^{t_{f}} (\vec{F}_{df,i} \cdot \vec{v}_{i} + \vec{F}_{gw,i} \cdot \vec{v}_{i}) dt \right] - E_{f}}{E_{0}-E_{f}}  \right|,
\end{equation}
where $E_{0}$ and $E_{f}$ are the initial and final energies, and the two 
terms in the sum under the integral are the work done by dynamical friction 
and gravitational radiation during the current stage of the code.
\begin{figure}
\includegraphics[width=240pt,height=240pt]{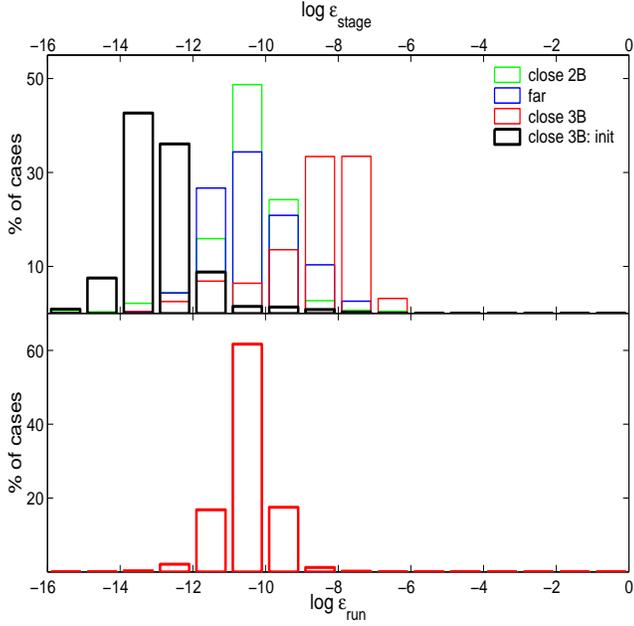}
\caption{Energy errors.  {\it Upper panel:} Errors given by equation~\eqref{epsE}, for each code stage:  close triple encounters, integrated with {\it Chain} (red); unperturbed binary evolution with the RK4 integrator (blue); and very close two-body encounters computed with {\it Chain} after the binary has coalesced (green).  The black (heavy) histogram shows the energy errors for close triple encounters normalized to the {\it initial} energy instead of the energy dissipated. {\it Lower panel:} Effective energy errors for the entire run, computed from equation~\eqref{epsErun}.}
\end{figure}
In the upper panel we separately plot 
the errors for close 3-body encounters, RK4 integration of the unperturbed
binary motion (``far''), and close two-body encounters computed with {\it Chain}
during the unperturbed binary evolution.  
The plot includes all code stages where the energy dissipated was at least 
$10^{-3}$ in code units, or about a part in $10^{5-6}$ of the initial 
binding energy of the system.   
The black (heavy) histogram shows
the errors for close encounters normalized to the {\it initial} energy 
instead of the dissipated energy in the denominator of equation~\eqref{epsE}, since the energy dissipated was very small in many close encounters.  In 
the lower panel we combine the energy errors from the various code stages to
get an effective energy error for each entire run,
\begin{equation}\label{epsErun} 
\epsilon_{run}=\frac{\sqrt{\epsilon_{1}^{2} \Delta E_{1}^{2}+\epsilon_{2}^{2} \Delta E_{2}^{2}+...+\epsilon_{n}^{2} \Delta E_{n}^{2}}}{\Delta E_{1}+\Delta E_{2}+...+\Delta E_{n}}.
\end{equation}
We had to combine the separate errors to obtain $\epsilon_{run}$ since the 
galactic potential is handled slightly differently during different stages 
of the code, e.g. the triaxial modification is applied only during the
RK4 integration.  In a large majority of cases $\epsilon_{run}$ falls 
between $10^{-12}$ and $10^{-9}$, and energy is conserved to better than a 
part in $10^{4}$ in every run.  The excellent energy conservation gives us 
confidence in the robustness of our integration methods.

\section{Results}

\subsection{Outcome statistics}

We begin with an overview of the outcomes of our 3-body simulations.  In 
subsequent sections we focus on various effects in more detail.  Our data
consists of eight sets of 1005 runs, each sampling a different distribution
of the ICs.  A set of 1005 runs took anywhere from 
$\sim$4 to $\sim$30 hours to finish on five 2.0 GHz Opteron processors, 
depending on the ICs. 

In our canonical runs (CN), we chose $\Delta_{m} = 0.3$ for the threshold 
merger mass ratio, modelled the stellar bulge as a Hernquist ($\eta = 2$) 
profile, started off the inner binary at $a_{hard}$, and generated the ICs
from a $5 \times 10^{13}$ M$_{\odot}$ halo at $z=0$.  In 
each of the remaining runs we varied one of these assumptions.  Runs D1 and 
D5 used $\Delta_{m} = 0.1$ and $\Delta_{m} = 0.5$ to explore the effects of 
widening or narrowing the range of BH mass ratios.  In runs MX we 
assigned a mass $m_{bh}[{\rm max}(M_{21},M_{22}),z_{f2}]$ instead of 
$m_{bh}(M_{21}+M_{22},z_{f2})$ to the intruding BH, as discussed in
\S2.1.  In runs BA and SA we started off the inner binary at $3a_{hard}$
and $a_{hard}/3$ instead of at $a_{hard}$.  We initialized the stellar 
bulge to an $\eta=1.5$ profile in runs SC, to explore the effect of a 
steeper inner cusp.  Finally in runs H1 we generated the ICs from a 
$1 \times 10^{13}$ M$_{\odot}$ halo at $z=0$, for total BH masses of 
$\sim 5 \times 10^{7}$ M$_{\odot}$, about an order of magnitude lower than
in our canonical runs.  Table 1 summarizes the outcomes.
\begin{table}
\caption{Summary of outcomes for eight different distributions of the ICs.  See text for explanation of the entries.}
\begin{tabular}{|p{0.6in}|p{0.15in}|p{0.15in}|p{0.15in}|p{0.15in}|p{0.15in}|p{0.15in}|p{0.15in}|p{0.15in}|p{0.15in}|}
\hline
Outcome & CN & D1 & D5 & MX & BA & SA & SC & H1 \\ \hline
Coalescence & & & & & & & &\\ \hline
One pair & 87\% & 68\% & 89\% & 84\% & 84\% & 95\% & 84\% & 75\% \\
Two pairs & 15\% & 13\% & 18\% & 16\% & 22\% & 13\% & 16\% & 10\% \\ \hline
Escape & & & & & & & &\\ \hline
Single & 15\% & 18\% & 17\% & 19\% & 21\% & 14\% & 14\% & 22\% \\
Double & 0.0\% & 0.0\% & 0.1\% & 0.0\% & 0.1\% & 0.2\% & 0.1\% & 0.3\% \\
Wander & 37\% & 51\% & 38\% & 42\% & 42\% & 26\% & 45\% & 50\% \\ \hline
Final state & & & & & & & &\\ \hline
Binary & 60\% & 67\% & 54\% & 57\% & 51\% & 65\% & 56\% & 63\% \\
Single & 38\% & 31\% & 45\% & 41\% & 47\% & 33\% & 43\% & 35\% \\
No BH & 1.3\% & 2.0\% & 1.2\% & 1.5\% & 1.4\% & 1.4\% & 1.2\% & 1.8\% \\ \hline
Core & & & & & & & &\\ \hline
$\langle M_{def} \rangle$ & 1.41 & 1.20 & 1.50 & 1.38 & 1.31 & 1.96 & 1.42 & 1.61 \\
$\Delta M_{def}$ & 0.48 & 0.37 & 0.57 & 0.44 & 0.52 & 0.44 & 0.57 & 0.61 \\ \hline
Termination & & & & & & & &\\ \hline
Sing escape & 14\% & 16\% & 16\% & 18\% & 18\% & 14\% & 12\% & 21\% \\
New binary & 61\% & 44\% & 61\% & 55\% & 53\% & 73\% & 54\% & 49\% \\
T.O. far & 22\% & 30\% & 22\% & 23\% & 20\% & 12\% & 31\% & 28\% \\
T.O. {\it Chain} & 2.1\% & 10\% & 1.7\% & 4.4\% & 8.4\% & 0.3\% & 1.6\% & 1.5\% \\
Crashed & 0.5\% & 0.1\% & 0.3\% & 0.3\% & 0.7\% & 0.2\% & 0.6\% & 0.1\% \\ \hline
\end{tabular}
\label{outcomest}
\end{table}

The first two rows give the percentage of cases in which (i) one BH pair 
coalescenced by the end of the run (i.e. by the time since the last 
major merger), and (ii) the remaining two BHs also coalesced within the run 
time.  At least one pair coalesced in a large majority of the runs for each
set of ICs that we tried.  The new system formed from the third BH and 
binary remnant also coalesced in $\sim$10-20\% of the cases.  Since we 
assume that stellar hardening of the new binary shuts off at $a_{hard}$, it 
can only coalesce by gravitational radiation from a highly eccentric orbit; 
we will discuss this topic further in \S3.2 and \S3.4.  The coalescence 
rate is somewhat lower ($\sim$ 68\%) in set D1, since (a) the hardening 
effect of the third body is lessened for more extreme mass ratios, and 
(b) mergers with mass ratios as low as $\Delta_{m}=0.1$ are more 
frequent, so the run time is typically shorter.  Naturally the coalescence 
rate is somewhat higher (95\%) in runs SA, where we begin with a tighter 
binary ($a_{0}=a_{hard}/3$, $\tau_{gw} \sim 10^{11-12}$ yrs).  Coalescence
is also significantly less common in runs H1.  This can be understood in 
light of equation~\eqref{gap} in \S1.1.  The separation between the scale 
set by the stellar kinematics ($a_{hard}$) and that set by gravitational
radiation ($a_{gw}$) is proportional to $m_{bin}^{-1/4}$.  Hence in lower
mass systems, coalescence is less likely relative to escape.  This 
observation motivates future study of triple BH dynamics in much lighter
systems.

The next three rows of the table give the fraction of runs in which (i) the 
single BH escaped the stellar bulge $+$ halo potential, (ii) all BHs (both 
the single and the binary) escaped the halo, and (iii) the single BH either 
escaped or remained wandering far out in the halo at the end of the run.  
The single escaped in $\sim$15-20\% of the runs in all cases.  If we also 
count runs where it remained wandering through the halo for of order a 
Hubble time, this fraction increases to $\sim$40\%.  Double escapes (of 
both the binary and the single BH) were very rare.  We get more wandering 
BHs in set D1, since a larger fraction of the released energy is apportioned 
to the escaper when it is relatively lighter, the dynamical friction time is
longer, and the run time is shorter.  Runs SA produced less 
wandering BHs since the binary pair more often coalesced before the intruder 
had a chance to harvest much of its energy.  Wandering was also more common
in set H1, due to the $m_{bh}^{-1/4}$ scaling discussed in the previous 
paragraph.  The escape fraction of course depends on the depth of
the galactic potential well.  Given the uncertainty and scatter in the 
$m_{bh}-M_{halo}$ relation and specificity of the prescription adopted,
we must expect these numbers to vary somewhat in studies with different
halo or stellar density models.

The entries under ``final state'' tell whether, at $z=0$, the 
galactic center hosts (i) a stalled BH binary, (ii) a single BH, or 
(iii) no BHs (neither the single nor the binary has yet returned to the 
center by dynamical friction).  About 50-70\% of the runs ended with a 
binary at the galactic center whose gravitational radiation time exceeded 
the time until $z=0$.  This includes cases where (a) the single was ejected 
to large distance and the binary settled to the center before it hardened 
enough to coalesce by gravitational radiation, (b) when the inner binary 
coalesced during a close encounter the outer binary coalescence time 
exceeded the remaining run time, or (c) the single and binary remnant both 
returned to the center after an ejection and formed a bound pair with a long
gravitational radiation time.  In most of the remaining cases
(30-50\%) the run ended with a single BH at the galactic center, or a binary
bound to coalesce before $t_{0}$.  This occurred when (a) the single was 
ejected to large distance and the binary (or remnant) settled to the center 
after having hardened to the point of coalescence through some combination 
of repeated interactions with the third BH and stellar dissipation, or 
(b) a new binary with a short gravitational radiation time formed following 
return from an ejection or coalescence during a close encounter.  In only a 
small fraction (1-2\%) of cases the run ended with the center empty of 
BHs.  Note also that this happened most often in runs where the last merger 
occurred recently, so the total {\it time} spent with the center empty of 
BHs is smaller still.

The next two entries give the mean and standard deviation of the core 
``mass deficit'' scoured out by the triple system, in units of the total BH 
mass $m_{bh}$.  For a galaxy modelled as an $\eta$-model with stellar mass 
parameter $M_{s}$, bulge scale radius $a_{s}$ and a break to inner slope 
$\gamma$ at $r_{b}$, we define the mass deficit $M_{def}$ by
\begin{multline}\label{mdef}
M_{def} = 4\pi \left[ \int_{0}^{r_{b}} \rho_{\eta}(r) r^{2} dr - 
\int_{0}^{r_{b}} \rho(r) r^{2} dr \right] = \\
M_{s} \left[(\frac{r}{r+a_{s}})^{\eta_{s}}-(\frac{r_{b}}{r_{b}+a_{s}})^{\eta_{s}} \right]-\frac{4\pi \rho_{bs}}{3-\gamma} r_{b}^{3} + {\rm D.M.},
\end{multline}
where $\rho_{bs}$ is the stellar density at $r_{b}$ and D.M. denotes the
corresponding dark matter terms.  The mass deficits are highly scattered 
within each set of runs, with typical $M_{def}/m_{bh} \approx 1.4 \pm 0.5$.
More extreme mass ratios (runs D1) tended to produce smaller cores, while
a narrower mass range (runs D5) gave somewhat larger ones.  The fraction of 
runs ending with very high mass deficits varied strongly with $\Delta_{m}$; 
for instance 17\% of cases ended with $M_{def}/m_{bh} > 2$ for 
$\Delta_{m}=0.5$, vs. only (11\%, 4.4\%) for $\Delta_{m}=$(0.3, 0.1).  The
large cores in set SA arose mostly from enhanced core scouring during the
creation of the initial hard binary, and so are more a consequence of the
ICs than of the triple interactions themselves.  This sensitivity of 
$M_{def}$ to the binary stalling radius is an interesting point in its 
own right.  The larger cores in runs H1 probably arise from the higher mean 
number of ejections and smaller fraction of runs ending in immediate 
coalescence as the BH mass is decreased.  21\% of the runs in this set ended
with $M_{def}/m_{bh} > 2$ and 8.7\% ended with $M_{def}/m_{bh} > 2.5$.  
The subject of core scouring will be discussed in further detail in \S3.5.

Both the core scouring effect and the coalescence rate induced by the 
encounters are significantly reduced for the extreme mass ratios in set D1.
One must keep in mind, however, that halo mergers with these mass ratios 
are much more frequent than those above $\Delta_{m}=0.3$ (see Fig. 3), so 
the ${\it cumulative}$ effect of these events may be as high or higher than
that of encounters with near-equal masses.  To quantify this statement our 
simulations would need to be embedded in a merger tree that follows the
formation of triple systems.

Finally, the last five lines in Table 1 give the statistics of the 
condition which formally terminated the run: (i) the single escaped the 
halo and the binary (or remnant) settled to rest at the galaxy center; 
(ii) the escaper and binary remnant formed a stalled binary or
coalesced (in a few percent of these cases a bound binary never actually
formed; the pair coalesced suddenly upon a very close periapsis passage from
an unbound orbit); (iii) the maximum physical time $t_{max}=t_{0}-t_{f0}$ 
was reached (in these cases one or more BHs were left wandering through the 
halo at the end of the run); (iv) the maximum time for a close encounter 
($t_{chn}=3 \times 10^{7}$ yrs for our canonical runs) was exceeded; or 
(v) the timestep went to zero or a limit on the number of timesteps was 
reached at some stage of the integration, which always occurred in $<$1\% of
cases.  Runs terminating on condition (iv) or (v) were left 
out when computing the upper entries in the table.   
    
In this slew of runs we have varied only a few of the relevant parameters; 
one might also try, for instance, varying or adding scatter to the halo mass
prescription, further steepening the stellar bulge profiles or adding a 
disk component, and exploring vastly different BH mass scales, in 
particular the much lower ($\sim 10^{4-5}$ M$_{\odot}$) masses that may be 
relevant at high redshift.  One of the advantages of our method is the 
relative ease of varying the model and ICs.  This paper should be viewed as 
a work in progress, in which we have developed a method that can be applied 
to 3-BH systems in whatever context they may arise.  Given the 
qualitative similarity of the outcomes in the runs we've performed so far, 
we will focus on the canonical (CN) runs in the more detailed presentation 
of our results.

\subsection{Efficient binary coalescence}

The inner binary begins at $a_{hard}$, where the gravitational radiation 
time is $\tau_{gw} \sim 10^{13-15}$ yrs, in our canonical runs.  It must 
shrink by a factor of $\sim$10 before gravitational radiation can cause 
coalescence in a Hubble time, or by a factor of $\sim$40 for $\tau_{gw}$ to 
become comparable to the dynamical friction time.  The intruder helps to 
bridge this gap in several ways: (a) direct hardening of the binary through 
repeated 3-body interactions, (b) enhanced stellar hardening by scattering 
of stars into the loss cone and motion of the binary about the 
nucleus, and (c) enhanced gravitational radiation 
losses due to thermalization of the eccentricity during the chaotic 
encounters and eccentricity growth via the Kozai resonance.  

We find that the combination of these mechanisms leads to coalescence of 
the inner binary within the time $t_{0}-t_{f0}$ between the merger that
formed the triple system and $z=0$ in a large majority of the runs.  It is
intructive to distinguish the systems that coalesce by ``collision'' during
a close 3-body encounter from those that gradually harden enough to
coalesce within the time $t_{0}-t_{f0}$, through the cumulative effect of
repeated encounters and loss-cone refilling while the binary wanders about
the nucleus.  In runs CN, 23\% of the systems undergo collision during the
first encounter, and another 19\% coalesce during later close encounters, 
for a net 42\% collision rate.  Thus about half of the total coalescence 
efficiency arises from collisions during close encounters, and the other
half comes from gradual hardening over the course of the simulation.  Kozai
oscillations account for most of the collisions during the first encounter,
while in later encounters random eccentricity variations are more likely to
result in coalescence, since the binaries are harder.  Our numbers
are reasonable based on analytic estimates of the collision rate
in chaotic encounters (e.g. \citealt{VK06}, chapter 11).

Fig. 8 shows the distribution of binary coalescence times.
\begin{figure}
\includegraphics[width=240pt,height=220pt]{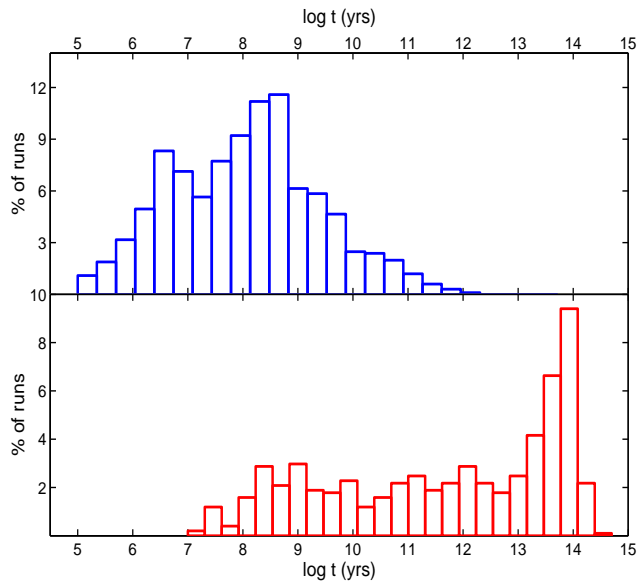}
\caption{Distribution of coalescence times for the inner binary ({\it upper panel}) and new binary ({\it lower panel}), formed by the third BH and coalesced binary remnant.  The lower plot includes only runs where a new binary formed, and not e.g. cases where the single escaped or was left wandering far from the galactic center at the end of the run.}
\end{figure}
The upper panel is for the inner binary, while the lower
panel is for the new system formed by the third BH and binary remnant.
In cases where coalescence occurred during the run, we plot the 
coalescence time recorded by the code.  In other cases we plot 
$t_{run}+t_{gr, end}$, where
$t_{run}$ is the total run time and $t_{gr, end}$ is the time
obtained by integrating the \citet{P64} equations from the state at the end 
of the run to coalescence.  The lower plot includes only those runs where 
the third BH ended up bound to the binary remnant, excluding, for instance,
cases where the single escaped the galaxy. In $\sim$15\% of the runs the 
new binary also coalesced within the time $t_{0}-t_{f0}$.  

Under circumstances where the gap-crossing mechanisms discussed in \S1.1 
fail, the efficient coalescence in massive triple systems provides a 
``last resort'' solution to the final parsec problem.

\subsection{The 3-body interactions}

Though the close encounters take up only a small fraction of the physical
time in our runs, it is the energy exchanges during these encounters that
determine the large-scale BH dynamics.  We now take a closer look at the 
3-body dynamics in a few representative cases.

In $\sim$20\% of the runs the binary swiftly coalesces during the first
encounter, usually with the help of the Kozai resonance.  Two examples of
this are shown in Fig. 9.  
\begin{figure}
\includegraphics[width=250pt,height=225pt]{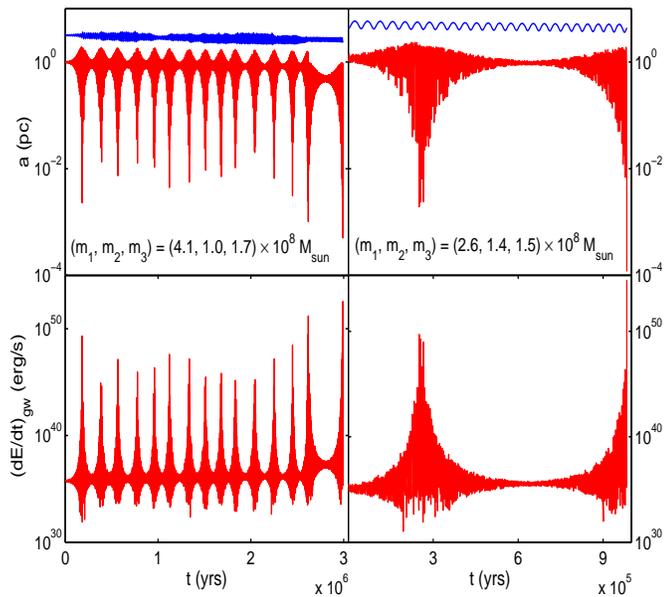}
\caption{Two examples of rapid coalescence by Kozai oscillations.  {\it Upper
panels:} Time evolution of the inner (red, lower) and outer (blue, higher)
binary separations.  {\it Lower panels:} Total gravitational radiation power,
averaged over Bulirsch-Stoer timesteps.  $m_{1}$ and $m_{2}$ are the masses
of the binary members and $m_{3}$ is the mass of the intruder.}
\end{figure}
The time evolution of the inner and outer binary 
separations is plotted for two different runs in the upper panels.  For a
circular orbit the separation would be roughly constant over an orbital 
period, or just a horizontal line in the figure.  On the left the inner
binary undergoes many Kozai oscillation cycles before coalescing.  Observe
that at the second-to-last eccentricity maximum, though it does not coalesce,
the binary radiates away a large amount of energy and passes through the 
next eccentricity minimum with a significantly reduced semi-major axis.  In
the example on the right, the binary coalesces after just one full Kozai
cycle.  The lower panels show the time evolution of the total gravitational
radiation power, averaged over the Bulirsch-Stoer timesteps.  Since the 
system starts on the verge of chaotic interactions where the
outer to inner binary semi-major axis ratio is small (so that the 
quadrupolar approximation breaks down), we get ``messy'' Kozai oscillations 
which can give way to catastrophic eccentricity growth at an unpredictable 
time.

Fig. 10 shows two examples of more complex runs.  The left panels summarize
the entire run, including all of the close encounters and ejections in 
between.  Each call to the {\it Chain} code or the unperturbed binary 
integrator is separated by dashed vertical lines.  The total time in each
stage is normalized to unity in order to see the full history of the run
at once, and not just distant ejections.  
The numbers on the plot are the actual times (in yrs) spent in each stage.  
\begin{figure}
\includegraphics[width=270pt,height=245pt]{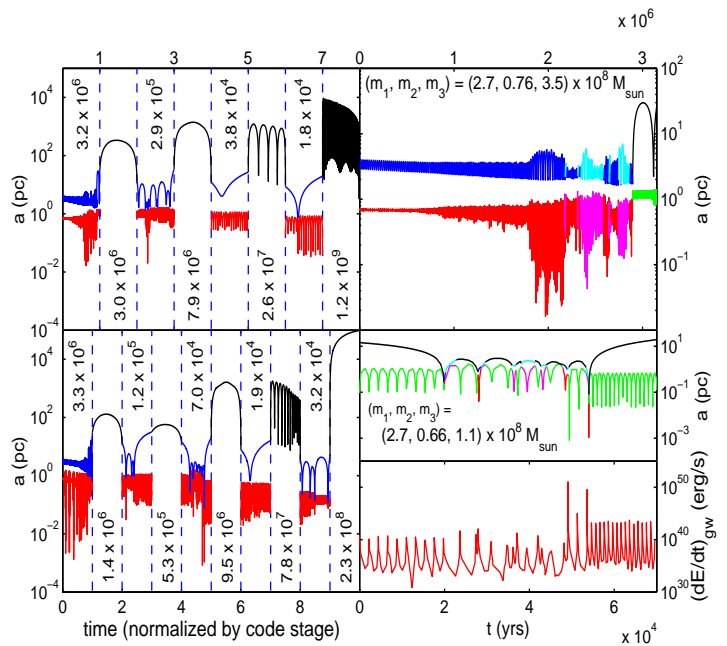}
\caption{Two examples of longer runs.  {\it Left panels:} Entire run, with time
spent in each code stage normalized to unity.  Actual times in years are
indicated by the numbers on the plot.  The red (inner binary) and blue
(outer binary) portions show the close encounters, while the black portions
show the calls to the RK4 integrator.  {\it Upper right:} Zoom
in on the first close encounter in the run at left.  Lines are color-coded
according to which pair is closest, to highlight the exchanges.  {\it Lower
right:} Zoom in on the third close encounter of the lower run, showing also
the total gravitational radiation power averaged over Bulirsch-Stoer
timesteps.  See text for further explanation of this figure.}
\end{figure}

Each run begins with a short period of secular evolution (illustrating the
remarkable stability of hierarchical triples even slightly within the 
Mardling-Aarseth boundary).  Dynamical friction brings the intruder in a bit
further to get chaotic interactions underway.  This can be seen more clearly
in the upper right panel, where we zoom in on the first close encounter at
left.  In this panel we also color-code the lines according to which two BHs
instantaneously form the closest pair, to show the numerous exchanges that
occur during close encounters.  Large-amplitude Kozai oscillations are 
present in the first encounter of the lower run, but no oscillations are 
seen in the upper run, where the initial inclination is below the critical
angle.   

After the first encounter, in both runs the outer components suffer a few
``near'' ($\sim$0.1-1 kpc) ejections before they get shot out to kpc scales
and come back by dynamical friction.  In the upper run, the single goes out
to $\sim$10 kpc, then comes back and forms a bound pair with the former 
binary, which has coalesced in the meantime.  The new binary is highly 
eccentric ($e$ $\sim$ 0.9998) and quickly coalesces by gravitational 
radiation.  In the lower run the single returns after the first kpc-scale 
ejection, strongly interacts with the binary one more time, and then 
escapes the galaxy.  The binary has a semi-major axis of 0.15 pc at the 
beginning of the final encounter, and its binding energy increases by 17\% 
in the interaction, imparting a velocity of $\sim$1400 ${\rm km/s}$ to the 
escaper.

In the lower right two panels we focus on the third encounter of this run,
which was selected because a significant amount of energy was lost to
gravitational radiation over its duration.  We can see that the radiation 
loss occurred during two very close approaches, by two different BH pairs.
If close 3-body encounters between BHs are sufficiently common in the 
universe, such gravitational radiation spikes could be detectable with LISA.

Fig. 11 shows the distribution of post-encounter velocities, for the single
and recoiling binary. 
\begin{figure}
\includegraphics[width=250pt,height=230pt]{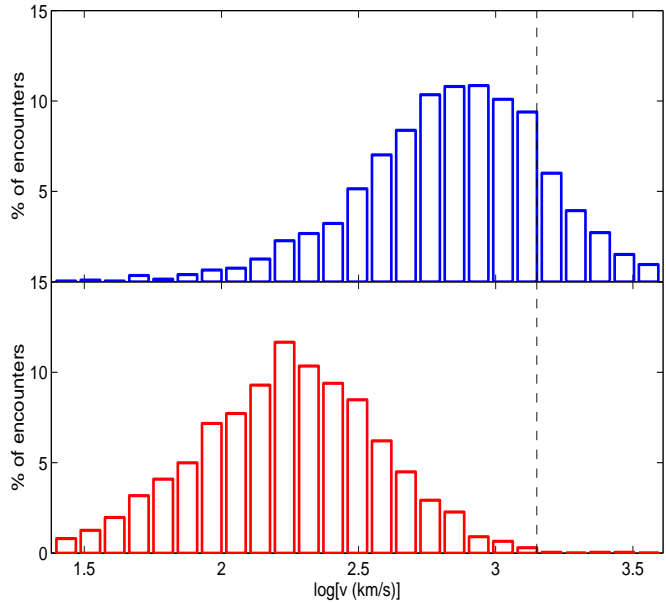}
\caption{Distribution of ejection velocities.  {\it Upper panel:} Single BH.  {\it Lower-panel:} Binary COM.  The total BH mass is typically $\sim 6 \times 10^{8}$ M$_{\odot}$.}
\end{figure}
Included in the plot are all close encounters in which (a) the
binary and single are unbound at the end of the encounter; (b) the binding
energy of the binary increases by at least 5\% (to avoid numerous 
``glancing'' encounters where $\delta F$ just barely exceeds the close 
encounter threshold), and (c) the encounter ends by $\delta F$ falling below
threshold (and not e.g. by coalescence of the binary or timing out).  
The dashed vertical line indicates the typical galactic (stellar bulge $+$ 
halo) escape velocity, $v_{esc} \sim 1400 \hspace{2pt} {\rm km/s}$.  We see 
that the single will sometimes escape the galaxy (or be ejected far out 
into the halo where the dynamical friction return time exceeds a Hubble 
time), but the binary will rarely go far.

The upper panel of Fig. 12 shows the distribution of fractional changes in 
the binding energy of the binary during close encounters, 
$1+\Delta = 1+(BE_{f}-BE_{0})/BE_{0}$.  The first encounter of each run is 
excluded from this plot, since it begins in a special hierarchical triple 
configuration and includes some dissipation by the dynamical friction used 
to bring in the intruder.  The red line shows the best fit to the form 
$f(1+\Delta)=K \Delta^{-1/2} (1+\Delta)^{-9/2}$, with the normalization $K$ 
depending on the mass ratios and intruder velocity, predicted by theory 
\citep{H75,VK06}.
\begin{figure}
\includegraphics[width=250pt,height=230pt]{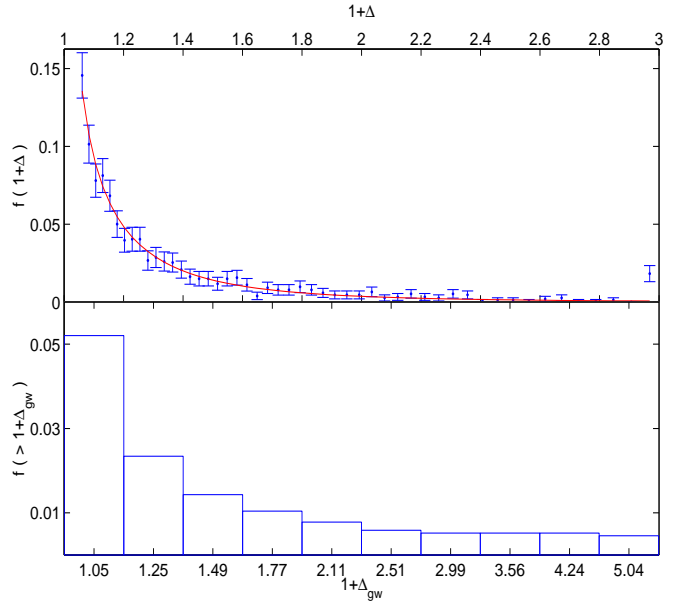}
\caption{Energy exchanges during close encounters.  {\it Upper panel:}  Distribution of the fractional change in the binding energy of the binary, $1+\Delta = 1+(BE_{f}-BE_{0})/BE_{0}$, in close encounters.  Blue points with 1.5$\sigma$ Poisson error bars are the simulation data; the red line shows a comparison with the shape predicted by theory (see text).  {\it Lower panel:} Fraction of encounters with the relative energy radiated as gravitational waves in the encounter greater than $1+\Delta_{gw} = 1+E_{gw}/BE_{0}$.  Both panels include only
encounters that end with the single BH unbound from the binary, and exclude the first encounter of each run, which starts in a special hierarchical triple configuration and also includes some dissipation by the dynamical friction used to bring in the intruder.}
\end{figure}
The lower panel shows the fraction of encounters with the relative energy radiated as gravitational waves in the encounter greater than $1+\Delta_{gw} = 1+E_{gw}/BE_{0}$.  This shows that gravitational radiation plays a 
significant role in the dynamics in only a few percent of the encounters 
ending in the escape of one component.  Another $\sim 20\%$ of the 
encounters end in coalescence; gravitational radiation of course plays a
significant role in all of these. 

Another point of interest is the statistics of the closest approach distances
between two-body pairs during the encounters.
Besides their intrinsic significance, the distances of closest approach 
are related to the extent of tidal stripping of the BHs during the 
encounters.  One can imagine that if some stars, or even the inner portion
of an accretion disk, remained bound to the individual BHs at the
end of an encounter, then some ejected SMBHs might become observable.

Since Bulirsch-Stoer timesteps are not at all infinitesimal (see \S2.4), 
we cannot simply take the minimum over the discrete timesteps to be the 
closest approach distance.  When the relative perturbation $\delta F$ from 
the third body is small, one can obtain the periapsis distance analytically 
in the Keplerian two-body approximation.  When $\delta F$ is larger, the 
minimum over the timesteps should give a better estimate since the 
timesteps tend to be smaller, but this statement is difficult to quantify.
To construct the distance of closest approach in our simulations, we first 
identify any step where $d|r|/dt=\hat{r} \cdot \vec{v}$ switches sign
from negative to positive and $|r| < 30000(r_{s1}+r_{s2})$ for any pair as
a ``passing step'' containing a close approach.  Here 
$\vec{r}=\vec{r}_{1}-\vec{r}_{2}$, $\vec{v}=\vec{v}_{1}-\vec{v}_{2}$, and
$r_{s1,2}$ are the Schwarzchild radii of the two pair members.  We then
iteratively bisect the timestep, evaluating $\hat{r} \cdot \vec{v}$ at each
bisection to find the place where it switches sign until the distance
between the two bodies converges to within a part in $10^{6}$.

Fig. 13 shows the distribution of the tidal radius $r_{tid}$ at closest 
approach, for the closest and second closest pair.
Since we are interested in observing ejected SMBHs, we
only include encounters where the single escaped with a velocity above
940 ${\rm km/s}$, the typical velocity needed to reach the stellar scale
radius of $\sim$3 kpc in our galactic model. 
\begin{figure}
\includegraphics[width=250pt,height=230pt]{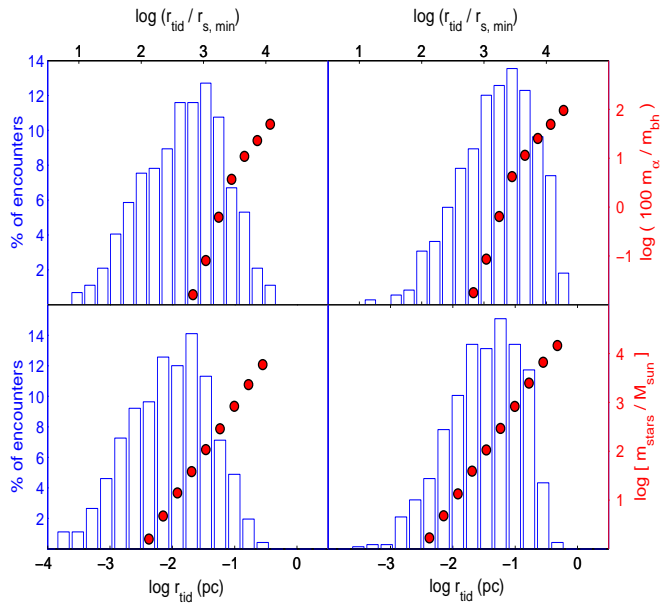}
\caption{Tidal radius $r_{tid}$ of the lighter pair member at closest approach.  In this plot we include only encounters where one BH escaped at a speed above 940 ${\rm km/s}$.  {\it Upper panels:} $r_{tid}$ in units of the Schwarzchild radius of the lighter pair member, $r_{s, min}$.  For reference, the red circles show the percent of the BH's mass contained within $r_{tid}$ in an $\alpha$-disk if the BH is accreting at the Eddington rate $\dot{m}_{Edd}$, averaged over the encounters in each bin.  {\it Lower panels:} $r_{tid}$ in pc.  Red circles show the mass in stars (in $M_{\odot}$) contained within $r_{tid}$ in the Hernquist profile used to model the stellar component of the galaxy.  {\it Left panels:} Closest pair; {\it Right panels:} Second closest pair.}
\end{figure}
$r_{tid}$ is defined by the equation
\begin{equation}
\delta a_{tid} \equiv \frac{Gm_{2}}{(d-r_{tid})^{2}} - \frac{Gm_{2}}{d^{2}} = \frac{Gm_{1}}{r_{tid}^{2}},
\end{equation}
where $m_{1}$ is the reference mass being stripped (the smaller pair 
member), $m_{2}$ is the other point mass, 
and $d$ is the distance between $m_{1}$ and $m_{2}$.  We solve this 
polynomial equation for $r_{tid}$ exactly rather than Taylor expanding 
about $r_{tid}/d = 0$ to get the familiar expression 
$r = (m_{1}/2m_{2})^{1/3}$ for the tidal radius, since
$r_{tid}/d$ is not generally small at closest approach for the near-equal
mass problem at hand.  The upper panels
show $r_{tid}$ in units of the Schwarzchild radius of the smaller BH.  The 
red circles indicate the percent of this BH's mass contained within 
$r_{tid}$ in an $\alpha$-disk \citep{SS73,FKR02,Nnotes} accreting at the 
Eddington rate $\dot{m}_{Edd}$, assuming
$\alpha=0.1$ and a radiative efficiency of $\epsilon=0.1$.  The lower panels
give $r_{tid}$ in pc, and here the red circles show the mass in stars within
$r_{tid}$ in the Hernquist model representing the stellar bulge.

The $\alpha$-disk model assumes that the disk is not self-gravitating and
breaks down as $\dot{m} \rightarrow \dot{m}_{Edd}$, so the red circles in the
upper panels are merely to give the reader an idea of the bound mass scales
associated with the approaches.  The tidal approximation is a pessimistic
estimate of the extent of the stripping since swift, one-time 
close passages would be impulsive (\citealt{BT87}, chapter 7).  We record 
only the single closest approach, so we cannot distinguish between such 
impulsive events and approaches that are part of periodic patterns in the 
trajectories. 

In a significant fraction of cases $r_{tid} \gsim 10^{4} r_{s}$ encloses a
substantial fraction of the BH's mass in accreting gas, so near-Eddington
accretion could continue for a duration of order the \citet{S64} time after
the slingshot ejection \citep{Loren}.  The enclosed stellar mass shown in
the lower panels is never nearly comparable to the BH mass, but in most
cases the escaper would drag some stars.  In principle one can imagine one
of these stars entering a giant phase and overflowing its Roche lobe,
producing detectable accretion onto the SMBH long after its ejection from 
the galactic center (e.g. \citealt{H04,K07}).  

\subsection{Distant Evolution and Binary Re-formation}

Slingshot ejections in triple encounters produce a population of
``wandering'' SMBHs in the halos of galaxies and intergalactic space 
\citep{V03a,VP05}.  Fig. 14 shows the total time spent by the single BH 
(upper panel) and binary/remnant (lower panel) at various distances from 
the galactic center, averaged over all 1005 runs.
\begin{figure}
\includegraphics[width=230pt,height=210pt]{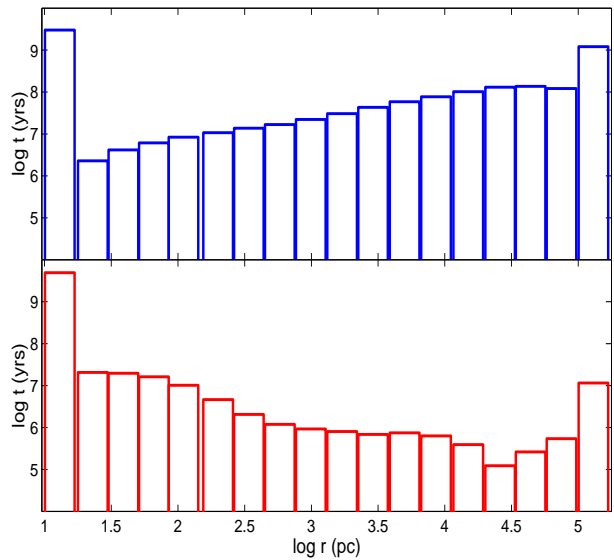}
\caption{Total time spent by the single BH ({\it upper panel}) and binary/remnant ({\it lower panel}) at various distances from the galactic center, averaged over all 1005 runs.  The lowest bin includes all distances below 17 pc and the highest bin includes all distances above $10^{5}$ pc.}
\end{figure}
The time spent in each distance bin is summed over the duration of each run,
and if the system reaches a steady state at time $t_{end} < t_{0}-t_{f0}$ 
then the state of the system after the run (until $z=0$) is included.  If a
component escapes, then the time $t_{0}-t_{end}$ is added to the highest bin;
if it settles to the center then this time is added to the lowest bin. 
The single is found wandering at large distances nearly half the time, 
while the binary spends the vast majority of its time at
the galactic center.  Over all of our runs, the total fraction of the time
spent with no SMBHs within 50 pc of the center since the formation of the 
halo hosting the triple system is only $\sim$ 1\%.  Hence we expect the 
ejections in triple-BH encounters at low redshift to produce very few 
nuclei empty of SMBHs.  A cD galaxy cluster, having hosted several dry 
mergers, might contain up to a few naked SMBHs wandering through the 
cluster halo as a result of single ejections.

The escaper remains wandering through the halo in only $\sim$40\%
of the runs.  In the other cases dynamical friction brings it back to the
center, where it becomes bound to the binary remnant and forms a new, hard
binary.  Fig. 15 shows the semi-major axis (upper panel) and eccentricity
(lower panel) distributions of the ``final state'' binaries in our 
simulations.  
\begin{figure}
\includegraphics[width=230pt,height=210pt]{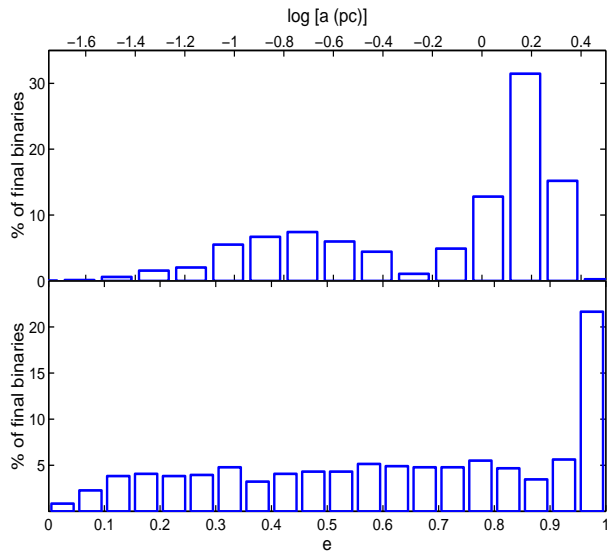}
\caption{Distribution of parameters of ``final state'' binaries.  The plot includes binaries that settled to the center after a single escape, and new binaries that formed from the single BH and inner binary remnant.  {\it Upper panel:} Semi-major axis.  {\it Lower panel:} Eccentricity.}
\end{figure}
This plot includes binaries formed when a pair coalesces
during a close encounter and is replaced by a single BH with its 
COM coordinates; cases where the original binary never coalesces,
but rather settles to the center and stalls after the single escapes; and
binaries that form from the third BH and coalesced remnant after ejections.

Whereas in the absence of triple
encounters we would expect most SMBHs to sit around $a_{hard}$, the 
encounters introduce a second population of stalled binaries at smaller
separations.  The eccentricities of the final binaries span the whole range
from 0 to 1.  Note the peak at very high eccentricity, arising mostly from
cases where the escaper rejoins the binary remnant from a radial orbit 
following a distant ejection, as in the run shown in the upper panel of Fig.
10.  Many of the binaries in this peak are expected to coalesce quickly by
gravitational radiation.  This
result has importance for LISA if 3-body ejections are common enough, 
since the gravitational radiation signature of a highly eccentric binary is
quite different from that of a circular binary.  An eccentric binary 
radiates at all integer harmonics of the orbital frequency, so its spectral
energy distribution peaks at higher frequencies, possibly enabling the 
detection of higher-mass SMBH binaries (e.g. \citealt{P01,EN06}).

However we caution the reader that the high-eccentricity coalescence rate
appears to be sensitive to the dynamical friction and core updating
prescriptions.  SMBH binary eccentricity evolution is a delicate
question to which simulators have obtained widely discrepant answers (e.g.
\citealt{A03,MM01,MMrev}).  While the conclusion that a high-eccentricity
population forms through distant ejections is robust, the question of 
whether these systems tend to coalesce or remain as an observable population
of stalled high-eccentricity binaries may depend on the details. 

\subsection{Core creation}

A number of studies have addressed the mark left on a stellar core by the
hardening of one or more BH binaries in independent succession 
\citep{MM01,R02,V03b,M06}.  To estimate the damage, consider a succession of
mergers $(M_{0},M'_{0}) \rightarrow M_{1}$, 
$(M_{1},M'_{1}) \rightarrow M_{2}$, ..., 
$(M_{N-1},M'_{N-1}) \rightarrow M_{N}$, between galaxies containing
BHs of mass $(m_{0},m'_{0})$, ... , $(m_{N-1},m'_{N-1})$, and having 
insufficient gas for significant stellar cusp regeneration.  Suppose that 
following each merger the BHs spiral in to $a_{hard}$ by dynamical 
friction on the stars, then cross the gap from $a_{hard}$ to $a_{gw}$ by 
some non-stellar mechanism, e.g. interaction with a modest amount of gas that
ends up in the nucleus through tidal torques associated with the merger.  
The total energy deposited in the stellar core is roughly
\begin{equation}\label{Edep}
E_{dep} \sim \sum_{i=0}^{N-1} \left( \frac{Gm_{i}m'_{i}}{a_{hard, i}} - \frac{Gm_{i}m'_{i}}{a_{inf, i}} \right),
\end{equation}    
where $a_{hard, i}$ is the hardening radius of the BH binary formed
in each merger, and the radius of influence $a_{inf, i}$ is the radius 
containing about twice the mass of the larger binary member in stars (e.g. 
\citealt{M06}).  Note that the right-hand side of~\eqref{Edep} is dominated 
by the first term in the parentheses, so the precise definition of 
$a_{inf, i}$ is not important.
If the inner density profile of a galaxy flattens from 
$d \ln \rho/d \ln r = 3-\eta$ to a shallower slope $\gamma$ within some 
core radius $r_{b}$, then we can define a core ``energy deficit'' by
\begin{equation}\label{Udef}
U_{def} = \pi \int_{0}^{\infty} \left[ \rho(r) \Phi(r) r^{2} - 
\rho_{\eta}(r) \Phi_{\eta}(r) r^{2} \right] dr,
\end{equation}
the difference between the binding energy of the galaxy with the density
break and that of the same galaxy, but with the density profile outside the
core extrapolated inward to the center.  In this equation $\rho =
\rho_{stars} + \rho_{halo}$ and $\Phi = \Phi_{stars} + \Phi_{halo}$ denote
the sums of the contributions to the density and gravitational potential
from the stellar and dark matter halo components.  The cross terms 
$\rho_{stars} \Phi_{halo}$ and $\rho_{halo} \Phi_{stars}$ contribute about 
10-20\% of the total binding energy, while the halo-halo terms are 
negligible.  We denote the outer slope by $3-\eta$ to match the \citet{T94} 
parameterization used in our galactic models.  We can estimate the size of 
the core created by the cumulative scouring action of the BH binaries 
formed in the succession of mergers by equating $U_{def}$ of 
equation~\eqref{Udef} to $E_{dep}$ given by equation~\eqref{Edep}.

The $U_{def}=E_{dep}$ prescription was introduced in order to estimate the 
extent of cusp destruction {\it before the binary hardens}.  If stalling 
were prevented by sufficient scattering of stars into the loss cone, an 
analogous energy argument would grossly overestimate the size of the core 
scoured out as the binary decayed from $a_{hard}$ to the separation where 
gravitational radiation could take over.  This is because a hard binary 
loses energy by ejecting stars at high velocities, often exceeding the escape
velocity of the entire galaxy.  Most of the energy released by the binary
goes into excess kinetic energy of these hyper-velocity stars rather than 
heating of the local medium.  Equations~\eqref{Edep} and~\eqref{Udef} capture
the essence of the core scouring in the limit of weak encounters (dynamical
friction), but for hard binaries we must view the cusp destruction as 
{\it mass ejection} rather than {\it energy injection}, once again following
the work of \citet{MV92} and \citet{Q96}.  A hard SMBH binary is defined by 
the fact that it hardens at a constant rate, $dE/dt$ $=$ const.
In the limit of very high orbital velocity ($w \gg \sigma$) this implies
that a constant mass in stars, comparable to the total BH mass, is
ejected from the galactic center per $e$-folding of the binary semi-major
axis,
\begin{equation}\label{jej}
\frac{1}{m_{bin}} \frac{dM_{ej}}{d\ln(1/a)} = \frac{1}{m_{bin}} 
\frac{dM_{ej}}{d\ln(E_{st})} \equiv J \approx 0.5,
\end{equation}
where $E_{st}$ is the energy transferred from the BH system to the 
stars.  We can estimate the core damage due to mass ejection by equating 
the total mass ejected by the binary to $M_{def}$ as defined in 
equation~\eqref{mdef}.

Now suppose that instead of coalescing without further damaging the stellar
core, the binary formed in the first merger in our sequence stalls at 
$a_{hard}$ until a third BH sinks in following the second merger.  On
the one hand, some energy that would have been injected into the stars as the
outer binary hardened may now instead be carried off as gravitational 
radiation or kinetic energy of a fast escaping BH, causing less 
damage to the stellar core than the decay of two separate binaries.  On the 
other hand, the intruder may continue scattering stars into the loss cone 
well after the inner binary reaches $a_{hard}$, and ejected BHs heat
the core by dynamical friction as their orbits pass repeatedly through the
dense nucleus \citep{BMQ04}.

To quantify these considerations our code evolves the core radius $r_{b}$ and
slope $\gamma$ along with the BH orbits to roughly account for the core
heating and mass ejection caused by the triple systems.
At the beginning of each run we initialize the core by injecting an energy
\begin{align}\label{Einit}
E_{init} = \frac{Gm_{1}m_{2}}{a_{hard,i}} &+
 \frac{G(m_{1}+m_{2})m_{3}}{a_{init,o}} \notag \\ 
- \frac{Gm_{1}m_{2}}{a_{inf,i}} &- \frac{G(m_{1}+m_{2})m_{3}}{a_{inf,o}}
\end{align}
into the parent $\eta$-model according to equation~\eqref{Udef}.  In runs 
where the inner binary starts at $a > a_{hard}$ we replace $a_{hard,i}$ with
$a_{init,i}$ in equation~\eqref{Einit}. In runs 
where it starts at $a$ $<$ $a_{hard}$ we also eject a mass 
$0.5m_{bin} \ln(a_{hard,i}/a_{init,i})$ according to equation~\eqref{mdef}
before the start of the run.

There is an ambiguity in the way we update the profile since energy may be
injected (or mass may be ejected) either by increasing $r_{b}$ to make the
core larger, or by decreasing $\gamma$ to make it shallower.  We resolved
this ambiguity by performing a rough fit to the $\gamma$ vs. 
$y \equiv M_{def}/m_{bh}$ data in \citet{M06}, to obtain the relation
\begin{equation}
\gamma \approx -0.0281y^{3} + 0.2451y^{2} - 0.7094y + 1.000
\end{equation} 
for $\eta = 2$, which gives sensible slopes for all $y \lsim 5$.  This
relation at least has the desired property that $\gamma \rightarrow 3-\eta$
as $M_{def} \rightarrow 0$, but the slope becomes quite shallow toward large
$M_{def}$.  The mass deficits are not sensitive to our prescription 
for $\gamma$. 
  
During the unperturbed binary integration, we increment the 
energy injected at each timestep $t \rightarrow t+\Delta t$ by
\begin{equation}\label{Einj}
\Delta E_{inj} = \sum_{i=1}^{2} \int_{t}^{t+\Delta t} (\vec{F}_{df,i} \cdot \vec{v}_{i}) \theta(r_{core} - r_{i}) dt,
\end{equation}
the work done on the single BH and binary COM by dynamical 
friction while the respective masses are located within a distance 
$r_{core} \equiv 1.5r_{b}$ of the galactic center.  When the binary is
located within $r_{core}$ and has not yet coalesced or settled to the center
and stalled, we also increment the ejected mass by
\begin{equation}\label{Mej}
\Delta M_{ej} = {\rm 0.5} m_{bin} \ln \left( \frac{E_{0}+\Delta E_{st}}{E_{0}} \right)
\end{equation}
at each timestep, where $E_{0}$ is the binding energy of the binary at the 
beginning of the timestep and $\Delta E_{st}$ is the change in binding 
energy due to stellar hardening during that step.  We cannot simply use the 
semi-major axis increment in equation~\eqref{Mej} since the binary may also
have hardened by gravitational radiation during the timestep.  At the 
beginning of each run, we also include the energy injected by dynamical
friction as the intruder spirals in before the onset of chaotic 
interactions.

Each time the total energy injected reaches 1\% of $E_{hard}$ or the total
mass injected reached 1\% of the total BH mass we update the core 
accordingly.  At the end of each run we record the final $M_{def}$, $r_{b}$, 
and $\gamma$.  Though both mass ejection and energy injection 
enter into our core growth algorithm, at the end both translate into a 
single effective $M_{def}$ as given by equation~\eqref{mdef}, which we
record for comparison with observed galaxies. 

Fig. 16 compares our calculated mass deficits to 14 cored, luminous 
elliptical galaxies with BH masses ranging from 
$\sim 10^{8}-3 \times 10^{9}$ $M_{\odot}$, with measured mass deficits.  
The upper panel shows data from our simulations.  The blue (left) histogram 
is the distribution at the beginning of the runs, reflecting the heating of 
the core by dynamical friction on the BHs as they sink into the initial 
configuration with the inner binary at $a_{hard}$, equation~\eqref{Einit}.  
This distribution also approximately represents the core damage expected 
for a single merger in which stellar hardening ceases at $a_{hard}$.  Note 
that a significant core ($M_{def}/m_{bh} \sim 0.5$) is scoured out even 
before the binary hardens and begins ejecting stars.  The middle 
(black dashed) histogram shows the core predicted for a series of two dry 
mergers, in both of which stellar hardening stops at $a_{hard}$.  The red 
(right) histogram is the distribution of cores at the end of our runs, 
reflecting the core scouring effect of the 3-body interactions.
\begin{figure}
\includegraphics[width=240pt,height=215pt]{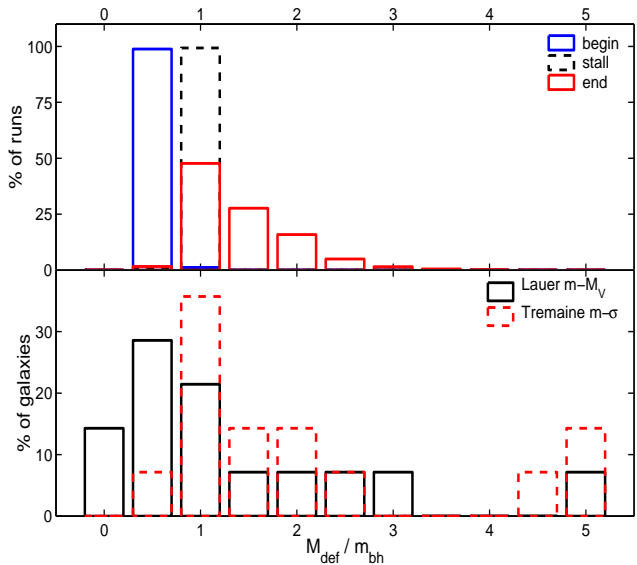}
\caption{Comparison of calculated and observed mass deficits.  {\it Upper panel:} Calculated mass deficits in units of the total BH mass, $m_{1}+m_{2}+m_{3}$.
Blue (left): Beginning of simulation, based on core heating during inspiral to the initial configuration.  Black dashed (middle): Net effect of two dry mergers in which core scouring stops at $a_{hard}$.  Red (right): End of simulation, reflecting net energy injection and mass ejection caused by close triple encounters.  {\it Lower panel:} Observed mass deficits in units of the SMBH mass \citep{G04,F06}.  In the red dashed histogram we determine the BH masses from the $m-\sigma$ relation of \citet{T02}, for all entries.  In the black solid histogram we instead use the $m-M_{V}$ relation of \citet{L06a} for those galaxies with $M_{V} < -22$, and the dynamically measured BH masses in the four cases where they are available.}
\end{figure}
The mass deficits in the lower panel are obtained in \citet{G04} and 
\citet{F06} by fitting the outer nuclear density profile to a S\'{e}rsic 
law \citep{S68}, and then subtracting a power-law fit to the inner core 
from the inward extrapolation of the S\'{e}rsic profile.  
The red dashed histogram in the lower panel shows the observed 
$M_{def}/m_{bh}$ with $m_{bh}$ computed from the $m_{bh}-\sigma$ relation of 
\citet{T02}.  However \citet{L06a} argue that luminosity may be a better 
predictor of BH mass than $\sigma$ for the most luminous elliptical 
galaxies ($M_{V} \lsim -22$), since their recent merger histories consisted 
mostly of passive (dissipationless) mergers in which both the BH mass and 
luminosity are simply additive. The $m_{bh}-\sigma$ relation is thought to 
arise from self-regulation of accretion onto the SMBH in gaseous mergers 
\citep{Silk,WL03b}, which does not apply in this context.  \citet{L06a} 
also show that an extrapolation of the $m_{bh}-L$ relation to the highest 
luminosities is more consistent with the observed $r_{core}-m_{bh}$ 
relation and provides a better match between the $z=0$ SMBH space density 
and the quasar population seen at $z \sim 2$ for reasonable quasar duty 
cycles.  In the black histogram, we used the observed BH masses for the 
four cases with dynamical mass measurements
\citep{G03,B98,M97,H94}.
\begin{table}
\caption{Mass deficits in galaxies with dynamically measured SMBH masses.  References: \citealt{G03} (G03), \citealt{B98} (B98), \citealt{M97} (M97), and \citealt{H94} (H94).}
\begin{tabular}{|c|c|c|c|c|}
\hline
Galaxy & $m_{bh}/M_{\odot}$ & $M_{V}$ & $M_{def}/m_{bh}$ & Reference\\ \hline
NGC 4291 & $3.1 \times 10^{8}$ & -20.64 & 1.8 & G03 \\
NGC 4374 & $1.6 \times 10^{9}$ & -22.28 & 1.4 & B98 \\
NGC 4486 & $3.0 \times 10^{9}$ & -22.71 & 2.9 & M97, H94 \\
NGC 4649 & $2.0 \times 10^{9}$ & -22.51 & 1.1 & G03 \\ \hline
\end{tabular}
\label{obsMbh}
\end{table}
For the rest of the galaxies we used the \citet{L06a} 
$m_{bh}-M_{V}$ relation to estimate the BH masses in galaxies with
$M_{V} < -22$, and the \citet{T02} $m_{bh}-\sigma$ relation for those with 
$M_{V} > -22$.  Showing both plots gives an idea of how much the mass 
deficits vary with $m_{bh}$ estimator.

In set CN $\sim$11\% of the runs resulted in cores with 
$M_{def}/m_{bh} > 2$.  In some rare runs where the binary was ejected to a
large distance and then brought back by dynamical friction, we obtained 
even higher mass deficits (up to $M_{def}/m_{bh} = 3-4$).  The binary is
more efficient at core scouring than the single BH since it is more 
massive.  If each independent binary inspiral adds $\sim 0.5m_{bh}$ to the
mass deficit, then the mean $M_{def}$ enhancement of $\sim 0.5m_{bh}$ due 
to the triple encounters is equivalent to one extra merger in the system's
history.  An $M_{def}$ one standard deviation above the mean is equivalent
to two extra mergers.  

\section{Discussion and Conclusions}

Triple-SMBH systems in galactic nuclei produce a range of phenomena and
signatures rather different from those expected if no more than two SMBHs
occupy them at a time.  We have developed an efficient numerical method for
following the evolution of 3-body systems in the centers of galaxies,
and used it to explore the outcomes of such encounters in massive
elliptical galaxies at low redshift.

We find a high efficiency of SMBH coalescence due to the encounters,
providing a ``last resort'' solution to the final parsec problem.  There
is, however, a caveat in extending this result immediately to all BH
masses.  If we define $a_{esc}$ to be the binary semi-major axis where
escape of one BH first becomes likely ($Gm_{bin}/\beta a_{esc}=v_{esc}^{2}$
where $\beta$ is a factor of order 10 for a Hernquist profile), then since
$v_{esc} \propto \sigma$ we have $a_{esc} \propto m_{bin}/\sigma^{2}
\propto m_{bin}^{1/2}$ if $m_{bin}$ obeys the $m_{bh}-\sigma$ relation 
$m_{bin} \propto \sigma^{4}$, so $a_{esc}/a_{gw} \propto m_{bin}^{-1/4}$.  
In other words, at smaller BH masses, the lightest BH is more likely to
escape the galaxy before driving the binary to coalescence by gravitational
radiation.  By focusing on massive galaxies we have chosen the systems
where the binary is {\it least} likely to coalesce by other means (e.g. gas
or massive perturbers), and {\it most} likely to coalesce in the next
merger with the help of 3-BH interactions.  We may address the efficiency
of triple-induced coalescence in much smaller-mass systems in future
studies.  

We find that close triple encounters can produce a population of
high-eccentricity binaries, whose gravitational radiation signal could
potentially be observable by LISA.  Such signals originate from Kozai
oscillations in hierarchical triples at high initial inclinations and highly
eccentric binaries formed following distant ejections.  As the eccentricity 
increases, the radiation spectrum peaks at progressively higher harmonics 
of the fundamental frequency, approaching a nearly flat spectrum as 
$e \rightarrow 1$ \citep{P01,EN06}.  A circular $10^{8-9}$ M$_{\odot}$ BH 
binary remains below the band of frequencies ($\sim 10^{-4}-10^{-1}$ Hz) 
detectable by LISA throughout its inspiral, but the occurrence of 
high-eccentricity coalescences could extend LISA's sensitivity into this 
mass range, or lengthen the duration of its sensitivity to $\sim 10^{6-7}$ 
M$_{\odot}$ events.  A highly eccentric binary produces a ``spiky'' 
waveform that looks quite different from that of a circular system 
(see Fig. 7 in \citealt{P01}).  Gravitational radiation ``spikes'' at very
close approaches during chaotic 3-body interactions could also produce
radiation bursts detectable by LISA.  

If triple encounters are indeed limited to massive systems at low redshift,
then the importance of these considerations is limited by the expected
event rate in this mass range, assuming efficient coalescence.  This rate
is highly uncertain, ranging from $\sim 1/{\rm yr}$ \citep{S05} to $\sim
1/1000{\rm yrs}$ \citep{RW05} depending on the merger and BH population
model adopted.  If 3-BH systems occur in other contexts, e.g. IMBHs in
galactic nuclei or star clusters, then the phenomena we have discussed may
be observationally relevant even if the high-mass SMBH event rate is low.
A detailed look at the gravitational waveforms expected from 3-body
encounters and their expected detection rates is an interesting topic for
a future study.

The slingshot ejections in triple encounters produce a population of
``wandering'' SMBHs in and outside the halos of galaxies.  In systems 
that have undergone several major dry
mergers (e.g. cD galaxy systems), one might expect a few such ejected SMBHs
to be floating in the vicinity.  As of yet, no probable way of observing
these wandering BHs has been proposed\footnote{Gravitational lensing is
difficult to search for without knowing in advance the location of the
BHs.}.  In principle one can imagine a star bound to the ejected SMBH
entering a giant phase and overflowing its Roche lobe, producing some
accretion onto the SMBH and an observable flare.  Single ejections could
also in principle affect BH-bulge correlations such as the $m_{bh}-\sigma$
relation, but since it is the lightest BH that gets ejected this effect
would fall well within the observed scatter in the correlations for just
one or two ejection events.

Triple interactions in galactic nuclei can have a large effect on the
expected properties of stable SMBH binaries in the local universe.  While
many models of binary formation predict mostly circular binaries around
$a_{hard}$, 3-body encounters produce binaries at all eccentricities.  
They also create a population of stalled binaries at separations 
significantly smaller than $a_{hard}$ but still larger than $a_{gw}$, as 
does any partial gap-crossing mechanism.

Better measurements and statistics on the mass deficits in cored elliptical
galaxies may provide clues on the history of the nuclear SMBH activity in
these systems.  Triple BH encounters produce a highly scattered 
distribution of core sizes, with mass deficits up to $\sim 2 \times$ higher
than expected for successive binary coalescences.  The apparent peak
at mass deficits of $\sim 0.5-1$ times the nuclear BH mass
in observed cores may very tentatively hint that multiple-BH encounters are
not the norm in these systems.  This signature of binary or multiple-BH
activity is appealling because (a) its duty cycle is the lifetime of the
galaxy; (b) it is present whether binary pairs stall or coalesce; and (c)
it can be observed even in the complete absence of radiative activity, such
as disk accretion or jet production.  However the interpretation of galaxy
cores is complicated by multiple mergers, the possibility of partial
stellar cusp regeneration from traces of cold gas,
and observational complications such as projection effects in nonspherical
galaxies and optimizing the fitting/extrapolation algorithm to best
represent the mass deficit.  There is a need for theoretical studies on the 
cores produced by SMBH mergers in triaxial galaxies, since triaxility seems 
to be the most likely candidate for a gap-crossing mechanism in dry mergers 
between gas-poor, giant ellipticals.  Inferring the nuclear histories of 
galaxies from their observed core properties will likely be a topic of much 
interest in the future.

\section*{acknowledgements} We would like to thank Suvendra Dutta for 
technical help at the parallel computing center of the Institute for Theory
and Computation (ITC), and Sverre Aarseth and Seppo Mikkola for making their
N-body algorithms and codes available.  We are also grateful to Scott
Hughes, Michael Eracleous, Marta Volonteri, and Fred Rasio for useful
discussions. This research was supported in part by an FQXi grant and
Harvard university funds.

\label{lastpage}

\vspace{0.5in}

\begin{thebibliography}{}
\bibitem[Aarseth (2003)]{A03} Aarseth, S.J. 2003, Ap\&SS 285, 367

\bibitem[Barkana \& Loeb(2001)]{BL01} Barkana, R. \& Loeb, A. 2001,
PhR, 349, 125

\bibitem[Berczik \etalb(2006)]{B06} Berczik, P., Merritt, D., Spurzem, R.,
\& Bischof, H. 2006, ApJ, 642, 21

\bibitem[Begelman \etalb(1980)]{BBR80} Begelman, M.C., Blandford, R.D., \&
Rees, M.J. 1980, Nature, 287, 25

\bibitem[Binney \& Tremaine(1987)]{BT87} Binney, J. \& Tremaine, S. 1987,
Galactic Dynamics (Princeton, New Jersey: Princeton University Press)

\bibitem[Blaes \etalb(2002)]{BLS02} Blaes, O., Lee, M.H., \& Socrates, A. 
2002, ApJ, 578, 775

\bibitem[Blanchet et al.(2005)]{Will} Blanchet, L., Qusailah, M.~S.~S., \&
Will, C.~M.\ 2005, ApJ, 635, 508


\bibitem[Bower \etalb(1998)]{B98} Bower, G.C., \etal 1998, ApJL, 492, 111  

\bibitem[Boylan-Kolchin \etalb(2004)]{BMQ04} Boylan-Kolchin, M., Ma, C., \&
Quataert, E. 2004, ApJ, 613, 37

\bibitem[Bulirsch \& Stoer(1966)]{BS66} Bulirsch, R. \& Stoer, J. 1966,
NuMat, 8, 1

\bibitem[Bullock \etalb(2001)]{B01} Bullock, J.S., Kolatt, T.S., Sigad, Y.,
Somerville, R.S., Kravtsov, A.V., Klypin, A.A., Primack, J.R., \& Dekel, A. 
2001, MNRAS, 321, 559

\bibitem[Byrd \etalb(1987)]{BSV87} Byrd, G.G., Sundelius, B., \& 
Valtonen, M. 1987, A\&A, 171, 16

\bibitem[Carroll \etalb(1992)]{C92} Carroll, S.M., Press, W.H., \& 
Turner, E.L. 1992, ARA\&A, 30, 499

\bibitem[Centrella(2006)]{Cent} Centrella, J. M. 2006, to appear in
conf. proc. for the Sixth International LISA Symposium, AIP; preprint
astro-ph/0609172
     
\bibitem[Chandrasekhar(1943)]{C43} Chandrasekhar, S. 1943, ApJ, 97, 255


\bibitem[Cohn \& Kulsrud(1978)]{CK78} Cohn, H. \& Kulsrud, R.M. 1978, ApJ,
226, 1087

\bibitem[Colpi \etalb(1999)]{CMG99} Colpi, M., Mayer, L., \& Governato, F.
1999, ApJ, 525, 720



\bibitem[Damour \& Deruelle(1981)]{DD81} Damour, T., \& Deruelle, N. 1981,
PhLA, 87, 81

\bibitem[de Zeeuw \& Carollo(1996)]{dZC96} de Zeeuw, P.T. \& Carollo, C.M.
1996, MNRAS, 281, 1333


\bibitem[Eisenstein \& Hu(1998)]{EH98} Eisenstein, D.J. \& Hu, W. 1998, ApJ,
496, 605

\bibitem[Enoki \& Nagashima(2006)]{EN06} Enoki, N. \& Nagashima, M. 2006,
PThPh, submitted, arXiv:astroph/0609377



\bibitem[Erickcek \etalb(2006)]{EKB06} Erickcek, A.L., Kamionkowski, M., \&
Benson, A.J. 2006, MNRAS, in press, arXiv:astroph/0604281

\bibitem[Escala \etalb(2005)]{ES05} Escala, A., Larson, R.B., Coppi, P.S.,
\& Mardones, D. 2005, ApJ, 630, 152 

\bibitem[Escala \etalb(2004)]{ES04} Escala, A., Larson, R.B., Coppi, P.S.,
\& Mardones, D. 2004, ApJ, 607, 765 

\bibitem[Favata \etalb(2004)]{FHH04} Favata, M., Hughes, S.A., \& 
Holz, D.E. 2004, ApJ, 607, 5

\bibitem[Ferrarese(2002)]{F02} Ferrarese, L. 2002, ApJ, 578, 90

\bibitem[Ferrarese \etalb(2006)]{F06} Ferrarese, L., \etal 2006, ApJS, 
in press, arXiv:astroph/0602297

\bibitem[Ferrarese \& Merritt(2000)]{FM00} Ferrarese, L. \& Merritt, D. 2000,
ApJ, 539, L9

\bibitem[Frank, King, \& Raine(2002)]{FKR02} Frank, J., King, A., \& 
Raine, D. 2002, Accretion Power in Astrophysics, 3rd Edition 
(Cambridge: Cambridge University Press)

\bibitem[Frank \& Rees(1976)]{FR76} Frank, J. \& Rees, M.J. 1976, 176, 633


\bibitem[Gebhardt \etalb(2000)]{G00} Gebhardt, K., \etal 2000, ApJ, 539, L13

\bibitem[Gebhardt \etalb(2003)]{G03} Gebhardt, K., \etal 2003, ApJ, 583, 92

\bibitem[Graham(2004)]{G04} Graham, A.W. 2004, ApJ, 613, 33


\bibitem[Haiman(2004)]{Haiman} Haiman, Z. 2004, ApJ, 613, 36 


\bibitem[Harms \etalb(1994)]{H94} Harms, R.J., \etal 1994, ApJL, 435, 35


\bibitem[Heggie(1975)]{H75} Heggie, D.C. 1975, MNRAS, 173, 729

\bibitem[Hein\"{a}m\"{a}ki(2001)]{H01} Hein\"{a}m\"{a}ki, P. 2001, 
A\&A, 371, 795

\bibitem[Hernquist(1989)]{H89} Hernquist, L. 1989, Nature, 340, 687

\bibitem[Hernquist(1990)]{H90} Hernquist, L. 1990, ApJ, 356, 359

\bibitem[Hills \& Fullerton(1980)]{HF80} Hills. J.G. \& Fullerton, L.W. 1980,
AJ, 85, 1281

\bibitem[Hoffman \& Loeb(2006)]{Loren} Hoffman, L., \& Loeb, 
A. 2006, ApJL, 638, L75 

\bibitem[Holley-Bockelmann \etalb(2002)]{H02} Holley-Bockelmann, K., Mihos,
J.C., Sigurdson, S., Hernquist, L., \& Norman, C. 2002, ApJ, 567, 817

\bibitem[Holman \etalb(1997)]{H97} Holman, M., Touma, J., \& 
Tremaine, S. 1997, Nature, 386, 254



\bibitem[Hopman \etalb(2004)]{H04} Hopman, C., Portegies Zwart, S.F., \&
Alexander, T. 2004, ApJL, 604, 101

\bibitem[Iwasawa \etalb(2005)]{I05} Iwasawa, M., Funato, Y., \& Makino, J.
2005, arXiv:astroph/0511391

\bibitem[Kitayama \& Suto(1996)]{KS96} Kitayama, T. \& Suto, Y. 1996, ApJ, 
469, 480


\bibitem[Komossa \etalb(2003)]{K03b} Komossa, S., Burwitz, V., Guenther, H.,
Predehl, P., Kaastra, J.S., \& Ikebe, Y. 2003, ApJ, 582, 15


\bibitem[Kormendy \& Gebhardt(2001)]{KG01} Kormendy, J. \& Gebhardt, K. 2001,
in 20th Texas Symposium on Relativistic Astrophysics, ed. J.C. Wheeler \& H.
Martel (Melville: AIP), 363

\bibitem[Kozai(1962)]{K62} Kozai, Y. 1962, AJ, 67, 591

\bibitem[Kupi \etalb(2006)]{K06} Kupi, G., Amaro-Seoane, P., \& Spurzem, R.
2006, MNRAS, in press, arXiv:astroph/0602125

\bibitem[Kuranov \etalb(2007)]{K07} Kuranov, A.G., Popov, S.B., 
Postnov, K.A., Volonteri, M., \& Perna, R. 2007, MNRAS, in press, 
arXiv:astroph/0702525

\bibitem[Kustaanheimo \& Stiefel(1965)]{KS65} Kustaanheimo, P. \& Stiefel
1965, ApJ, 97, 255

\bibitem[Lacey \& Cole(1993)]{LC93} Lacey, C. \& Cole, S. 1993, MNRAS, 
262, 627

\bibitem[Lauer \etalb(2006)]{L06a} Lauer, T.R., \etal 2006, 
arXiv:astroph/0606739


\bibitem[Lee(1993)]{L93} Lee, M.H. 1993, ApJ, 418, 147  

\bibitem[Lightman \& Shapiro(1977)]{LS77} Lightman, A.P., \& Shapiro, S.L.
1977, ApJ, 211, 244

\bibitem[Macchetto \etalb(1997)]{M97} Macchetto, F., Marconi, A., Axon, D.J.,
Capetti, A., Sparks, W., \& Crane, P. 1997, ApJ, 489, 579

\bibitem[Magorrian \etalb(1998)]{M98} Magorrian, J., \etal 1998, AJ, 
115, 2285

\bibitem[Manrique \& Salavador-Sole(1996)]{MPS96} Manrique, A. \& 
Salvador-Sole, E. 1996, ApJ, 467, 504

\bibitem[Maoz \etalb(1995)]{M95} Maoz, D., Filippenko, A.V., Ho, L.C., 
Rix, H.-W., Bahcall, J.N., Schneider, D.P., \& Macchetto, F.D. 1995, ApJ, 
440, 91

\bibitem[Maoz \etalb(2005)]{M05} Maoz, D., Nagar, N.M., Falcke, H., \&
Wilson, A.S. 2005, ApJ, 625, 699

\bibitem[Marconi \& Hunt(2003)]{MH03} Marconi, A. \& Hunt, L.K. 2003, AJ, 
589, L21

\bibitem[Mardling \& Aarseth(2001)]{MA01} Mardling, R. \& Aarseth, S. 2001, 
MNRAS, 321, 398

\bibitem[Merritt(2006)]{M06} Merritt, D. 2006, arXiv:astroph/0603439

\bibitem[Merritt \etalb(2004)]{M04} Merritt, D., Milosavljevic, M., 
Favata, M., Hughes, S.A., \& Holz, D.E. 2004, ApJ, 607, 9

\bibitem[Merritt \& Milosavljevi{\'c}(2005)]{MMrev} Merritt, 
D., \& Milosavljevi{\'c}, M.\ 2005, Living Reviews in Relativity, 8, 8 


\bibitem[Merritt \& Poon(2004)]{MP04} Merritt, D. \& Poon, M.Y. 2004, ApJ,
606, 788

\bibitem[Merritt \& Quinlan(1998)]{MQ98} Merritt, D. \& Quinlan, G.D., ApJ,
498, 625

\bibitem[Mikkola \& Aarseth(1990)]{MA90} Mikkola, S. \& Aarseth, S. 1990,
CeMDA, 47, 375

\bibitem[Mikkola \& Aarseth(1993)]{MA93} Mikkola, S. \& Aarseth, S. 1993,
CeMDA, 57, 439 

\bibitem[Mikkola \& Valtonen(1990)]{MV90} Mikkola, S. \& Valtonen, M.J. 1990,
ApJ, 348, 412

\bibitem[Mikkola \& Valtonen(1992)]{MV92} Mikkola, S. \& Valtonen, M.J. 1992,
MNRAS, 259, 119

\bibitem[Milosavljevic \& Merritt(2001)]{MM01} Milosavljevic, M. \& Merritt,
D. 2001, ApJ, 563, 34




\bibitem[Narayan(2003)]{Nnotes} Narayan, R., Theory of Thin Accretion 
Disks, course notes distributed in Astronomy 219, Spring 2003, Harvard 
University

\bibitem[Navarro \etalb(1997)]{NFW} Navarro, J.F., Frenk, C.S., \& 
White, S.D.M 1997, ApJ, 490, 493

\bibitem[Peng \etalb(2005)]{P06} Peng, C.Y., Impey, C.D., Ho, L.C., 
Barton, E.J., \& Rix, H. 2006, ApJ, 640, 114

\bibitem[Perets \etalb(2006)]{PHA06} Perets, H.B., Hopman, C., \& 
Alexander, T. 2006, ApJ, submitted, arXiv:astroph/0606443

\bibitem[Peters(1964)]{P64} Peters, P.C. 1964, PhRv, 136, 1224

\bibitem[Pierro \etalb(2001)]{P01} Pierro, V., Pinto, I.M., Spallicci, A.D.,
Laserra, E., \& Recano, F. 2001, MNRAS, 325, 358



\bibitem[Quinlan(1996)]{Q96} Quinlan, G.D. 1996, NewA, 1, 35

\bibitem[Quinlan \& Hernquist(1997)]{QH97} Quinlan, G.D., \& Hernquist, L.
1997, NewA, 2, 533 

\bibitem[Raig, Gonzalez-Casado, \& Salvador-Sole(2001)]{MPS01} Raig, A.,
Gonzalez-Casado, G., \& Salvador-Sole, E. 2001, MNRAS, 327, 939

\bibitem[Ravindranath \etalb(2002)]{R02} Ravindranath, S., Ho, C., \& 
Filippenko, A.V. 2002, ApJ, 566, 801

\bibitem[Rhook \& Wyithe(2005)]{RW05} Rhook, K.J., \& Wyithe, J.S.B., 
MNRAS, 361, 1145


\bibitem[Rodriguez \etalb(2006)]{R06} Rodriguez, C., Taylor, G.B., 
Zavala, R.T., Peck, A.B., Pollack, L.K., \& Romani, R.W. 2006, ApJ, 646, 49

\bibitem[Roos(1981)]{R81} Roos, N. 1981, A\&A, 104, 218


\bibitem[Salpeter(1964)]{S64} Salpeter, E.E. 1964, ApJ, 140, 796

\bibitem[Salvador-Sole, Solanes, \& Manrique(1998)]{MPS98} Salvador-Sole, E.,
Solanes, J.M., \& Manrique, A. 1998, ApJ, 499, 542

\bibitem[Sanders \etalb(1988)]{S88} Sanders, D.B., Soifer, B.T., 
Elias, J.H., Madore, B.F., Matthews, K., Neugebauer, G., \& Scoville, N.Z. 
1988, ApJ, 325, 74

\bibitem[Saslaw \etalb(1974)]{S74} Saslaw, W.C., Valtonen, M.J., \& 
Aarseth, S.J. 1974, ApJ, 190, 253

\bibitem[S\'{e}rsic(1968)]{S68} S\'{e}rsic, J.L. 1968, Cordoba, 
Argentina: Observatorio Astronomico, 1968

\bibitem[Sesana \etalb(2005)]{S05} Sesana, A., Haardt, F., Madau, P., \&
Volonteri, M. 2005, ApJ, 623, 23

\bibitem[Shakura \& Sunyaev(1973)]{SS73} Shakura, N.I, \& Sunyaev, R.A. 1973,
A\&A, 24, 337


\bibitem[Silk \& Rees(1998)]{Silk} Silk, J., \& Rees, M.J. 1998, A\&A, 331, 
L1 



\bibitem[Spergel \etalb(2006)]{S06} Spergel, D. \etal 2006, 
arXiv:astroph/0603449


\bibitem[Springel \etalb(2005)]{SDH05} Springel, V., Di Matteo, T., \& 
Hernquist, L. 2005, MNRAS, 361, 776

\bibitem[Stiefel \& Scheifele(1971)]{SS71} Stiefel, E.L. \& Scheifele, G.
1971, Linear and Regular Celestial Mechanics (New York: Springer-Verlag)


\bibitem[Taffoni \etalb(2003)]{TEA03} Taffoni, G., Mayer, L., Colpi, M., \&
Governato, F. 2003, MNRAS, 341, 434

\bibitem[Tremaine \etalb(1994)]{T94} Tremaine, S., Richstone, D.O., Byun, Y.,
Dressler, A., Faber, S.M., Grillmair, C., Kormendy, J., \& Lauer, T.R. 1994,
AJ, 107, 634

\bibitem[Tremaine \etalb(2002)]{T02} Tremaine, S., \etal 2002, ApJ, 574, 740



\bibitem[Valtonen(1976)]{V76} Valtonen, M.J. 1976, A\&A, 46, 435 

\bibitem[Valtonen \etalb(1994)]{V94} Valtonen, M.J., Mikkola, S., 
Heinamaki, P., \& Valtonen, H. 1994, ApJS, 95, 69 

\bibitem[Valtonen \& Karttunen(2006)]{VK06} Valtonen, M. \& Karttunen, H. 2006, The Three-Body Problem (Cambridge, UK: Cambridge University Press)

\bibitem[Volonteri \etalb(2003a)]{V03a} Volonteri, M., Haardt, F., \&
Madau, P. 2003, ApJ, 582, 559

\bibitem[Volonteri \etalb(2003b)]{V03b} Volonteri, M., Madau, P., \& 
Haardt, F. 2003, ApJ, 593, 661

\bibitem[Volonteri \& Perna(2005)]{VP05} Volonteri, M. \& Perna, R. 2005,
MNRAS, 358, 913


\bibitem[Wyithe \& Loeb(2003a)]{WL03a} Wyithe, J.S.B. \& Loeb, A. 2003, ApJ,
590, 691

\bibitem[Wyithe \& Loeb(2003b)]{WL03b} Wyithe, J.S.B. \& Loeb, A. 2003, ApJ,
595, 614

\bibitem[Wyithe \& Loeb(2005)]{WL05} Wyithe, J.S.B. \& Loeb, A. 2005, ApJ, 
634, 910 

\bibitem[Yu(2002)]{Y02} Yu, Q. 2002, MNRAS, 331, 935

\end{thebibliography}
\end{document}